\documentclass[aps,prx,twocolumn,superscriptaddress,preprintnumbers,floatfix,nofootinbib]{revtex4-2}

\usepackage{float}
\usepackage{nicefrac}
\usepackage{slashed}
\usepackage{mathtools}
\usepackage{amsfonts}
\usepackage{amssymb}
\usepackage{amsmath}
\usepackage{graphicx}
\usepackage{subfigure}
\usepackage{array}
\usepackage{dcolumn}
\usepackage[mathscr]{euscript}
\usepackage{bm}
\usepackage{latexsym}
\usepackage{longtable}
\usepackage{hyperref}
\usepackage{verbatim}
\usepackage{epsfig}
\usepackage{color}
\usepackage[capitalise]{cleveref}
\usepackage{scalerel}
\usepackage{feynmp-auto}
\usepackage{braket}
\usepackage{orcidlink}
\usepackage{booktabs}
\usepackage{enumitem}
\usepackage{dsfont}
\hypersetup{ 
    pdfnewwindow=true,
    colorlinks=true,
    allcolors=[RGB]{31 119 180}
} 

\renewcommand{\min}{{\text{min}}}

\newcommand{\bcol}{\left[ \begin{array}{c}}
\newcommand{\ecol}{\end{array} \right]}
\newcommand{\beq}{\begin{eqnarray}}
\newcommand{\eeq}{\end{eqnarray}}

\newcommand{\kev}{\ensuremath{{\mathrm{\,ke\kern -0.1em V}}}\xspace}
\newcommand{\mev}{\ensuremath{{\mathrm{\,Me\kern -0.1em V}}}\xspace}
\newcommand{\gev}{\ensuremath{{\mathrm{\,Ge\kern -0.1em V}}}\xspace}
\newcommand{\tev}{\ensuremath{{\mathrm{\,Te\kern -0.1em V}}}\xspace}

\begin{document}
\title{S-matrix informed neural networks for amplitude analysis}
\preprint{JLAB-THY-26-4916}
\newcommand{\catania}{INFN Sezione di Catania, Catania I-95123, Italy}
\newcommand{\ceem}{Center for  Exploration  of  Energy  and  Matter, Indiana  University, Bloomington,  Indiana  47403,  USA}
\newcommand{\indiana}{Department of Physics, Indiana  University, Bloomington,  Indiana 47405,  USA}
\newcommand{\jlab}{Theory Center, Thomas  Jefferson  National  Accelerator  Facility, Newport  News,  Virginia  23606,  USA}
\newcommand{\lbnl}{Nuclear Science Division, Lawrence Berkeley National Laboratory, Berkeley, California 94720, USA}
\newcommand{\messina}{Dipartimento di Scienze Matematiche e Informatiche, Scienze Fisiche e Scienze della Terra, Universit\`a degli Studi di Messina, Messina I-98166, Italy}
\newcommand{\odu}{Department of Physics, Old Dominion University, Norfolk, Virginia 23529, USA}
\newcommand{\wm}{Department of Physics, The College of  William \& Mary, Williamsburg, Virginia 23187, USA}
\newcommand{\ucb}{Department of Physics, University of California, Berkeley, California 94720, USA}
\newcommand{\uned}{Departamento de F\'isica Interdisciplinar, Universidad Nacional de Educaci\'on a Distancia (UNED), Madrid E-28040, Spain}

\author{Wyatt~A.~Smith\,\orcidlink{0009-0001-3244-6889}}\email{wyattsmith@lbl.gov}\affiliation{\lbnl}\affiliation{\ucb}\affiliation{\wm}\affiliation{\messina}
\author{Arkaitz~Rodas\,\orcidlink{0000-0003-2702-5286}}\email{arodasbi@odu.edu}
\affiliation{\jlab}\affiliation{\odu}
\author{Marius~D.~Thomas\,\orcidlink{0009-0005-1945-594X}}\affiliation{\ucb}
\author{C\'esar~Fern\'andez-Ram\'irez\,\orcidlink{0000-0001-8979-5660}}\email{cefera@ccia.uned.es}\affiliation{\uned}
\author{Giorgio~Foti\orcidlink{0009-0000-9791-3823}}\affiliation{\messina}\affiliation{\catania}
\author{Lin~\surname{Qiu}\orcidlink{0000-0002-2683-8851}}\affiliation{\jlab}\affiliation{\odu}
\author{Adam~P.~\surname{Szczepaniak}\orcidlink{0000-0002-4156-5492}}\affiliation{\jlab}\affiliation{\ceem}\affiliation{\indiana}
\author{Alessandro~Pilloni\orcidlink{0000-0003-4257-0928}}\affiliation{\messina}\affiliation{\catania}

\collaboration{Joint Physics Analysis Center}

\date{\today}

\begin{abstract}
Reconstructing scattering amplitudes from finite, noisy, and mutually inconsistent measurements is an ill-posed inverse problem common to many reactions relevant to particle physics. We introduce S-matrix informed neural networks (SINNs), and demonstrate their ability to learn scattering amplitudes directly from data while respecting first principles. We further develop a novel data selection procedure, which uses the response of constrained neural network ensembles to identify a set of experiments compatible with first principles, and with each other. We apply this framework to $\pi\pi$ scattering, producing reusable amplitudes and correlated uncertainties without relying on a fixed functional form. We validate our results against residual model dependencies and training biases through closure tests and ablations. We find negligible impact of model architecture on our results. Our workflow unifies physics-constrained representation learning, data selection, and uncertainty quantification. Our strategy is transferable to other scattering processes, and other constrained physics problems limited by inconsistent data.
\end{abstract}

\maketitle

\section{Introduction}
\label{sec:intro}

Understanding how hadrons interact is a prerequisite for New Physics searches at the Intensity Frontier. The scattering amplitude encodes this information at the quantum level. It provides access to the excited spectrum of quantum chromodynamics (QCD), i.e. resonance masses and widths, hadronic contributions to precision observables like the muon $g-2$~\cite{Lees:2012cj,KLOE-2:2017fda,Colangelo:2018mtw,Aoyama:2020ynm}, and final-state interactions that enter $CP$-violation measurements~\cite{Colangelo:2017fiz,LHCb:2019jta,Garrote:2022uub}. For any of these quantities to be trustworthy, the amplitude must be inferred accurately and uncertainties must be estimated faithfully.

Unlike high-energy processes, where perturbative QCD gives direct access to dynamics, low-energy strong-interaction analyses are data-driven because there is no nonperturbative analytic solution of QCD. The amplitude has to be reconstructed from measurements. That reconstruction always requires some functional representation, flexible enough to follow the data and still respecting the scattering matrix (S-matrix) principles of unitarity, analyticity, and crossing symmetry~\cite{Eden:1966dnq}. The standard toolbox of parametrizations is broad, running from Breit-Wigner and $K$-matrix forms to unitarized chiral amplitudes, $N/D$, conformal and Pad\'e expansions, continued-fraction continuations, and dispersively constrained fits~\cite{Dobado:1989qm,Dobado:1996ps,Nieves:1999bx,Oller:1997ti,Oller:1998hw,Oller:1998zr,GomezNicola:2001as,Pelaez:2004xp,Kaminski:2002pe,Yndurain:2007qm,GarciaMartin:2011jx,Binosi:2022ydc}. Necessarily, each such functional form spans only a limited region of the space of admissible amplitudes, and the precision each claims is always conditional on the chosen representation, on the quality of the experimental input, and on the extrapolation prescription used to determine resonance parameters and subthreshold quantities~\cite{JPAC:2021rxu,Hanhart:2026sjg,Shastry:2026vlf,Shastry:2026mkk}. The resulting model and data-selection dependence is a tightly coupled source of epistemic and stochastic uncertainties that propagate into every derived quantity, especially for the determination of broad light scalars~\cite{Morgan:1990ct,Hanhart:2008mx,Hanhart:2014ssa,Masjuan:2014psa,Caprini:2008fc,Caprini:2016uxy,Rodas:2023nec}.  In particular, tensions among the various data sets that feed these systematics are often left uncontrolled. Global data sets routinely disagree beyond their quoted uncertainties (see, e.g., Refs.~\cite{BaBar:2012bdw,KLOE:2010qei}). This leaves a hard inverse problem with three competing demands: use as much of the available data as possible, constrain the space of amplitudes no more than the physics requires, and quantify uncertainty despite the coupling between data selection and model dependencies. The problem of reconstructing a continuous amplitude from a finite set of noisy, partly inconsistent measurements is ill-posed: the data alone do not determine a unique amplitude, and every analysis resolves the remaining degeneracy through some form of regularization, whether stated or hidden in an \textit{ansatz}. Traditional parameterizations are not well suited to this type of demanding analysis. They are seldom flexible enough to fit a full, conflicting data set without \textit{ad hoc} decisions about the validity of individual measurements, and, because they are rigid, a small change in the assumed form can shift the extracted mass, width, or coupling of a resonance with no physical reason for the change. There is no first-principles functional form to reach for.

In contrast, a neural network is ideal for such analyses; a suitable physics-driven neural representation can supply real-axis amplitudes without choosing a fixed \textit{ansatz}. Removing the explicit parametrization lets the data set the shape of the amplitude, in place of a functional prior. Casting the full analysis as a learning problem also makes the pipeline systematically testable: the architecture, hyperparameters, data-selection criteria, and even the physics input, become explicit and reproducible choices that can be varied and studied. Neural physics analyses have been made possible through physics informed neural networks~\cite{Raissi2019}, and have propagated to particle physics, though its reach has so far been limited. The NNPDF collaboration pioneered the idea of replacing an assumed parton-distribution parametrization with a neural network~\cite{Ball:2008by,NNPDF:2021njg}. Other approaches include the S-matrix neural bootstrap~\cite{Dersy:2023job,Gumus:2024lmj}, as well as lineshape-based pole classification via deep learning~\cite{Sombillo:2021yxe,Ng:2021ibr}, and simulation-based inference applied to the $\rho(770)$ pole under model misspecification~\cite{Sadasivan:2025kjj}.

In this work we develop a single framework that starts from the data, imposes the S-matrix constraints during optimization, automatically selects compatible data from an incompatible global data set, and delivers model-independent amplitudes. We will apply such a framework to $\pi\pi$ scattering, the natural first target and one of the few that can be studied rigorously.  It is the lightest stable hadronic system under the strong interaction and a primary source of hidden model dependences: pions dominate the hadronic final states that detectors record, so most hadronic measurements are filtered through $\pi\pi$ dynamics, and the same final-state interactions that complicate precision and New Physics searches are the ones encoded in the $\pi\pi$ amplitude. The measurement of $\pi\pi$ scattering is extremely difficult. The data are not measured directly but extracted from $\pi N \to \pi\pi N^\prime$, which injects its own model dependence before any analysis begins. At low energy the amplitude is tied to chiral perturbation theory~\cite{Colangelo:2001df,DescotesGenon:2001tn}; it contains the long-debated, very broad $\sigma/f_0(500)$~\cite{Pelaez:2015qba,ParticleDataGroup:2024cfk} alongside the narrow $f_0(980)$ perched on the $K\bar K$ threshold; and the global data set has been historically inconsistent~\cite{Kaminski:2002pe,Yndurain:2007qm,Ochs:2013gi,Perez:2015pea}. At the same time, several modern dispersive analyses provide reference values to compare against~\cite{GarciaMartin:2011cn,Moussallam:2011zg,Pelaez:2024uav,Pelaez:2026fdrpoles}. The $\pi\pi$ system is an ideal, yet unforgiving, stress test for a constrained, data-driven ML analysis.

We introduce S-matrix informed neural networks (SINNs) and apply them to obtain the low-energy $\pi\pi$ scattering amplitude and its uncertainty. We produce a direct amplitude representation, introduce a novel data selection procedure, and provide access to our final line shapes. \cref{sec:setup} describes the physical constraints each SINN must satisfy, providing a comprehensive presentation of the of the $\pi\pi$ scattering  process and its associated dynamics. We describe the available experimental data in \cref{sec:data}. \cref{sec:method} presents the architecture, multi-objective loss, curriculum, and ensemble construction; the data-selection procedure appears in~\cref{sec:selection}. \cref{sec:results} presents the partial-wave amplitudes, accuracy of physics constraints, and scattering lengths. We present model robustness and validation studies in~\cref{sec:validation}. Our conclusions are left to \cref{sec:discussion}. The Appendices give the numerical Roy and Regge implementation, supplementary partial-wave results, and all relevant training and validation details.

\section{Pion scattering}
\label{sec:setup}

Before describing the SINN, we define the scattering quantities and the principles that any admissible amplitude must satisfy. Throughout this paper we use the system of natural units $\hbar=c=1$.

\subsection{S-matrix constraints}
\label{sec:generic}
The S-matrix encodes a scattering process by relating its asymptotic in- and out-states of the interacting particles~\cite{Newton:1982,GoldbergerWatson:1964}. The $T$-matrix corresponds to the non-trivial scattering process, defined by \mbox{$\mathrm{S} = \mathds{1}+ i T$}. The strong interaction is diagonal in total isospin, and the $\pi\pi$ matrix element defines three different amplitudes with \mbox{$I=0,1,2$} as
\begin{align}\label{eq:isospin-amplitude}
&{}_I\langle \pi(p_4)\pi(p_3)|\, T \,|
\pi(p_1)\pi(p_2)\rangle_{I'} \nonumber\\
&\qquad =
(2\pi)^4\delta^{(4)}(P_f-P_i) \delta_{II'}
 \mathcal{A}^I(s,t,u) ,
\end{align}
with the kinematics entirely determined by the  Mandelstam invariants 
\begin{equation}
s=(p_1+p_2)^2,\qquad
t=(p_1-p_3)^2,\qquad
u=(p_1-p_4)^2,
\end{equation}
fulfilling $s+t+u=4m_\pi^2$.

We neglect isospin-breaking effects, so the amplitude depends on $I$ but not on its projection $I_z$. For the \mbox{$s$-channel} amplitudes, we decompose into partial waves via
\begin{align}
\mathcal{A}^I(s,t)= 32\pi \sum_{\ell=0}^{\infty}(2\ell+1)\,t_\ell^I(s)\,P_\ell(\cos\theta),
\label{eq:partial-wave-expansion}
\end{align}
where $\theta$ is the scattering angle in the center-of-mass frame. The inverse projection is
\begin{align}
t_\ell^I(s)=\frac{1}{64\pi}\int_{-1}^{1}dz_s\,P_\ell(z_s)\,\mathcal{A}^I(s,t(s,z_s)),
\label{eq:partial-wave-projection}
\end{align}
with \mbox{$z_s\equiv\cos\theta=\left( s+2t-4m_\pi^2\right)\slash \left( s - 4m_\pi^2\right)$}. We denote a partial wave in spectroscopic notation by $L_{I}$, where $L\in \{S, ~P, ~D, ~F, ~G\}$ corresponds to $\ell \in \{0,~1,~2,~3,~4\}$ respectively. Thus $S_0 \equiv t_0^0,~P_1 \equiv t_1^1$ and so forth. The first S-matrix principle relevant to the physical-region partial waves is {\it unitarity}. It follows from probability conservation, $\mathrm{S}^\dagger \mathrm{S}=\mathds{1}$, and for elastic $\pi\pi$ scattering implies
\begin{align}
\mathrm{Im}~t_\ell^I(s)=\frac{2q}{\sqrt{s}}\,|t_\ell^I(s)|^2,
\end{align}
where \mbox{$q=\sqrt{s-4m_\pi^2}\slash 2$} is the center-of-mass momentum. This relation couples the real and imaginary parts of each partial wave above threshold, fixes the discontinuity across the physical right-hand cut, and determines the connection between different Riemann sheets. It is implemented by writing the partial wave in terms of a phase shift $\delta_\ell^I(s)$, and an inelasticity $\eta^I_\ell(s)$,
\begin{align}
t_\ell^I(s)=\frac{\sqrt{s}}{2q}\frac{1}{2i}\left(\eta_\ell^I(s)e^{2i\delta_\ell^I(s)}-1\right).
\label{eq:partial-wave}
\end{align}
Below inelastic thresholds, {\it i.e.} when the system can only contain two-pion final states, $\eta_\ell^I(s)=1$. Above an inelastic threshold, $0\leq \eta^I_\ell \leq 1$; the inelasticity may vary to accommodate the initial state's coupling to additional channels.

The next principle is {\it analyticity}. Causality implies that scattering amplitudes can be analytically continued to complex Mandelstam invariants, with singularities of the amplitude fixed by physical intermediate states and thresholds. Consequentially, if we know the amplitude in the entire real-axis above threshold, we can in principle reconstruct the amplitude in the whole complex plane uniquely. On the first Riemann sheet, the amplitude contains bound-state poles, when present, as well as right-hand and left-hand branch cuts generated by direct- and crossed-channel particle thresholds. Complex poles on the first Riemann sheet are associated with tachyons. For each partial wave we define the scalar quantity $\mathrm{S}_\ell^I = 1 + 2i\rho(s) t_\ell^I(s),$ with $\rho(s) = 2q/\sqrt{s}$. In the elastic region, continuation through the $\pi\pi$ cut gives \mbox{$\mathrm{S}_{\ell,\mathrm{II}}^{I}(s)=1/\mathrm{S}_\ell^I(s)$} and \mbox{$t^I_{\ell, \text{II}}(s)=t_\ell^I(s)/\mathrm{S}_\ell^I(s)$}. A resonance pole on that sheet is therefore a zero of the first-sheet partial-wave S-matrix. Ignoring this structure when extracting resonance poles can produce large systematic errors, especially for broad states, or poles far from the physical region \cite{Caprini:2005zr,Caprini:2008fc,Pelaez:2015qba,Rodas:2023gma,Rodas:2023nec,ParticleDataGroup:2024cfk}. This provides a stringent global constraint on the allowed representations of the scattering amplitude, because analyticity extends to every isospin amplitude and every partial wave. Near particle thresholds, analyticity also gives rise to barrier factors that suppress the phase shifts as $q^{2\ell+1}$, so higher partial waves are suppressed at lower energies, and the low-energy dynamics is typically dominated by the $S$- and $P$-waves.

The final principle is {\it crossing symmetry}, which associates particles and antiparticles. Relativistic crossing implies that processes related by interchanging external particles are associated with the same analytic amplitude evaluated in different kinematic regions. The three $s$-channel isospin amplitudes can therefore be expressed in terms of a single scalar amplitude $\mathcal{A}(s,t,u)$,\footnote{Although kinematically one of the Mandelstam variables is redundant due to $s+t+u=4m_\pi^2$, expressing the amplitudes in terms of the three allows us to make crossing symmetry explicit.}
\begin{align}
\mathcal{A}^{I=0}(s,t,u) &= 3\,\mathcal{A}(s,t,u) + \mathcal{A}(t,s,u) + \mathcal{A}(u,t,s), \nonumber\\
\mathcal{A}^{I=1}(s,t,u) &= \mathcal{A}(t,s,u) - \mathcal{A}(u,t,s), \\
\mathcal{A}^{I=2}(s,t,u) &= \mathcal{A}(t,s,u) + \mathcal{A}(u,t,s), \nonumber
\end{align}
where $\mathcal{A}$ is written in the charged pion basis.
The crossing relations above show that the same analytic amplitude can be evaluated in the $s$-, $t$-, or \mbox{$u$-channel} physical regions, corresponding to the permutations $\mathcal{A}^{I}(s,t,u)$, $\mathcal{A}^{I}(t,s,u)$, and $\mathcal{A}^{I}(u,t,s)$. These channels are algebraically related by crossing, and unitarity requires the partial waves in each channel to have a \mbox{right-hand} cut starting at the two-pion threshold, $4m_\pi^2$. Consequently, when a \mbox{fixed-$t$} amplitude is viewed in the \mbox{$s$-channel}, $\mathcal{A}^{I}(s,t,u)$ has the usual right-hand \mbox{$s$-channel} cut, while the right-hand cut of the crossed $u$ channel appears as a left-hand cut starting at \mbox{$s=-t$}~\cite{Roy:1971tc,Basdevant:1972uv,Pennington:1973hs,Ananthanarayan:2000ht,Buettiker:2003pp,Hoferichter:2015dsa}.

In practice, experiments constrain only a finite number of partial waves in the physical region through the extraction of phase shifts and inelasticities. Unitarity can then be built directly into the partial-wave parametrization through \cref{eq:partial-wave}. Analyticity is more difficult and is only sometimes imposed in restricted energy regions or checked {\it post hoc}. Crossing symmetry is often neglected in phenomenological descriptions of experimental data or lattice QCD spectra. Dispersion relations provide the standard way to enforce these constraints simultaneously~\cite{Roy:1971tc,Ananthanarayan:2000ht,GarciaMartin:2011cn}.

\subsection{Roy equations}
\label{sec:pipi}

The Roy equations are a set of dispersion relations which provide the quantitative form in which analyticity, unitarity, and crossing symmetry enter in our learning problem~\cite{Roy:1971tc,Basdevant:1972uu,Basdevant:1972uv,Basdevant:1973ru,Pennington:1973hs,Pennington:1973xv,Ananthanarayan:2000ht,GarciaMartin:2011jx}. The construction starts from Cauchy's theorem, applied to \mbox{fixed-$t$} amplitudes at a value $s$ in the complex variable plane. Given the analytic structure presented above, one can relate the amplitude at a given point $s$ to integrals over its discontinuities across the cuts~\cite{Mandelstam_1962}. Unitarity identifies the physical right-cut discontinuity with the imaginary part of the amplitude. Crossing symmetry relates the \mbox{left-cut} discontinuity to the \mbox{right-cut} discontinuity over different isospin amplitude combinations. Since the \mbox{fixed-$t$} amplitudes may diverge at high energies, the dispersion relation must be `subtracted', meaning the high energy integral kernels must be suppressed by a polynomial in their denominator in exchange for a complementary polynomial with free parameter coefficients added to the dispersion relation. We will later see that these free parameters have a direct physical interpretation. The standard Roy construction uses two subtractions, and for real values of $s$ at fixed $t$ we have,
\begin{align}
&\mathrm{Re}\,\mathcal{A}^I(s,t) =P^I(s,t) \nonumber\\
&\quad +\sum_{I^\prime}\mathrm{P.V.}\!\int_{4m_\pi^2}^{\infty}ds^\prime\mathcal{K}^{II^\prime}(s,t;s^\prime)\,\mathrm{Im}\,\mathcal{A}^{I^\prime}(s^\prime,t)~.
\label{eq:fixed-t-disp}
\end{align}
Here $P^I$ is a first-degree polynomial, $\mathrm{P.V.}$ stands for Cauchy's principal value, and $\mathcal{K}^{II^\prime}$ denotes the suppressed amplitude dispersive kernels. Crossing symmetry fixes the structure of the subtraction polynomial, which for $\pi\pi$ scattering can be written in terms of the two scattering lengths, $a_0^0$ and $a_0^2$, associated with the isospin $0$ and isospin $2$ $S$ waves, respectively. The physical information contained within the integral enters through the discontinuity across the \mbox{right-hand} cut, written here as the imaginary part of the amplitude.

In practice, the \mbox{fixed-$t$} integral is split at a matching point $s_h$,
\begin{align}
\int_{4m_\pi^2}^{\infty} \left[ \dots \right] ds^\prime =
\int_{4m_\pi^2}^{s_h} \left[ \dots \right] ds^\prime
+\int_{s_h}^{\infty} \left[ \dots \right] ds^\prime .
\label{eq:roy-split}
\end{align}
The low-energy ($s<s_h$) part is written in terms of partial waves, which are the objects learned by the SINN. The high-energy ($s>s_h$) part is absorbed into a driving term supplied by a fixed Regge description, introduced in the next subsection. This matching point therefore separates the region where the fit is directly constrained by partial-wave data from the asymptotic region needed to complete the dispersion relation.

After projecting the fixed-$t$ dispersion relation, \cref{eq:fixed-t-disp}, with~\cref{eq:partial-wave-projection}, we obtain
\begin{equation}
\begin{aligned}
\mathrm{Re}\,\tilde{t}_\ell^I(s)
&=
\mathrm{ST}_\ell^I(s)
+ \mathrm{KT}_\ell^I(s)
+ \mathrm{DT}_\ell^I(s), \\
\mathrm{KT}_\ell^I(s)
&=
\sum_{\ell^\prime I^\prime}\mathrm{P.V.}\!\int_{4m_\pi^2}^{s_h}\!\! ds^\prime\,
K_{\ell\ell^\prime}^{II^\prime}(s,s^\prime)\,\mathrm{Im}\,t_{\ell^\prime}^{I^\prime}(s^\prime).
\end{aligned}
\label{eq:roy}
\end{equation}
This is the working form of the Roy equations used in our analysis. The tilde in $\tilde{t}_\ell^I(s)$ denotes the Roy-reconstructed partial wave. The subtraction term $\mathrm{ST}_\ell^I$ is the low-energy polynomial inherited from the twice-subtracted fixed-$t$ dispersion relation. The kernel term $\mathrm{KT}_\ell^I$ contains the low-energy partial-wave imaginary parts below $s_h$. The driving term $\mathrm{DT}_\ell^I$ contains the high-energy contribution above \mbox{$\sqrt{s_h}=1.5~\mathrm{GeV}$}, where the amplitude is represented by fixed Regge input and projected numerically. Typically for pion scattering, one imposes these equations on the three dominant partial waves, $S_0$, $P_1$, and $S_2$, though the kernels still include the higher-wave set needed to complete the dispersive input.

These Roy equations are rigorous only inside their domain of applicability. The fixed-$t$ dispersion relations and partial-wave projection lead to a finite region in the complex $s$ plane where the equations can be used as constraints; outside this region, additional assumptions about the amplitude and its high-energy behavior enter. On the real axis, we will impose the Roy residual only for \mbox{$2m_\pi < \sqrt{s}<1.1~\mathrm{GeV}$}, within the standard low-energy domain of validity~\cite{Roy:1971tc,Ananthanarayan:2000ht,Colangelo:2001df,Caprini:2005zr,GarciaMartin:2011cn}. Notice that Roy kernels mix isospin and angular momentum channels. The reconstructed real part of one wave depends on the imaginary parts of many other waves. This crossed-channel averaging lets each constrained output wave inherit information from the full amplitude, including the left-hand cut. On the real axis, the diagonal kernels require a principal-value prescription because they contain a pole at $s^\prime=s$. Details of the kernels, principal-value treatment, matching point, and Regge input are given in Appendices~\ref{app:roy_numerics} and~\ref{app:regge}.

For the three constrained channels, the subtraction terms are
\begin{subequations}
\begin{align}
\mathrm{ST}_0^0(s) &= \frac{a_0^0 + 5\, a_0^2}{3} + \frac{2\, a_0^0 - 5\, a_0^2}{12\, m_\pi^2}\, s, \\
\mathrm{ST}_1^1(s) &= \frac{2\, a_0^0 - 5\, a_0^2}{72\, m_\pi^2}\left(s-4\, m_\pi^2\right), \\
\mathrm{ST}_0^2(s) &= \frac{2\, a_0^0 + a_0^2}{6} - \frac{2\, a_0^0 - 5\, a_0^2}{24\, m_\pi^2}\, s.
\end{align}\label{eq:ST}
\end{subequations}
Here $a_0^0$ and $a_0^2$ are the scattering lengths in the $S_0$ and $S_2$ partial waves. The scattering lengths are therefore both physical observables and subtraction constants for the $\pi\pi$ system. 

Finally, Roy equations are one member of a broader family of possible dispersive constraints. Once-subtracted variants, often called Garc\'ia-Mart\'in--Kami\'nski--Pel\'aez--Yndur\'ain (GKPY) equations in the $\pi\pi$ literature, reduce the weight of the subtraction constants at higher energies~\cite{GarciaMartin:2011cn}, producing smaller errors for some observables. Hite-Steiner equations~\cite{Hite:1973pm} apply the same logic and extend the applicability region for processes where crossing relates different reactions, such as $\pi K$ and $\pi N$ scattering~\cite{Buettiker:2003pp,Hoferichter:2015hva}.

\begin{figure*}[t]
    \centering
    \includegraphics[width=\linewidth]{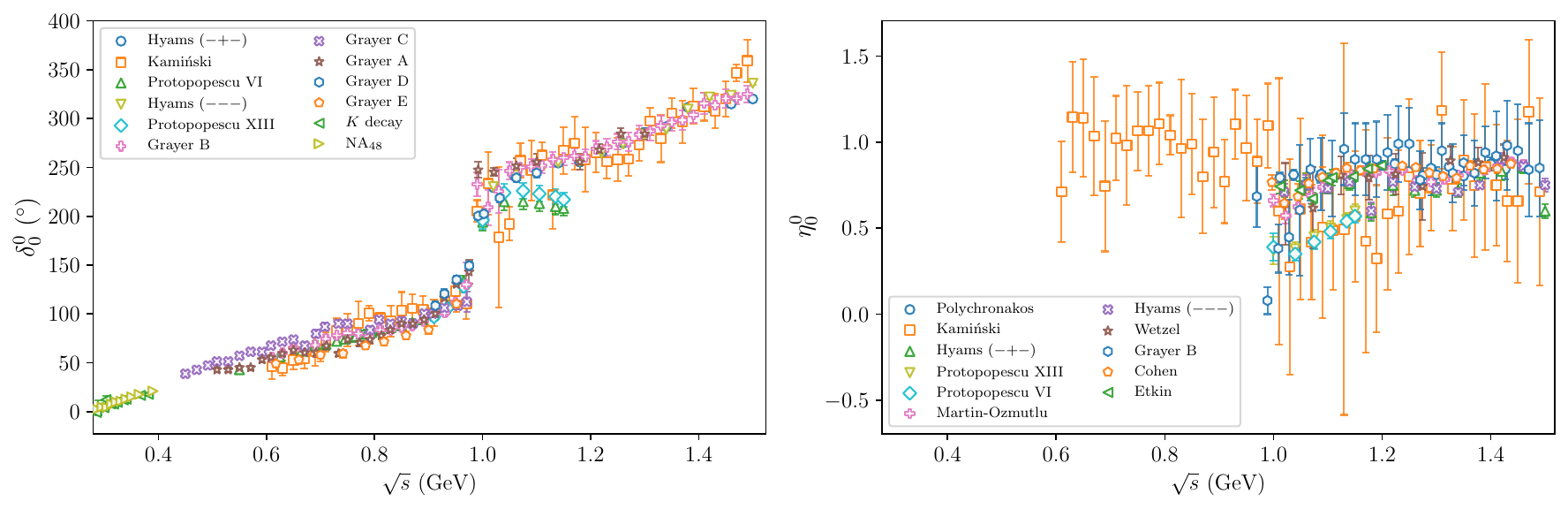}
    \caption{Experimental input for the $S_0$ wave before data selection. The figure includes all $S_0$ phase-shift and inelasticity data studied in this work. The spread between experiments illustrates the data tensions that motivate the selection analysis described in \cref{sec:selection}.}
    \label{fig:s0_data}
\end{figure*}

\subsection{High-energy constraints}
\label{sec:high_low_constraints}

The high-energy input in our work enters through the split in \cref{eq:roy-split}. Above $s_h$, a finite partial-wave expansion is no longer the proper description for the scattering amplitude. Many partial waves contribute comparably, so the truncated expansion does not converge. Equivalently, the intermediate resonances accumulate and overlap, producing an amplitude that becomes smooth as the energy increases. In this regime, the standard description comes from Regge theory: amplitudes are organized by $t$-channel exchanges lying on Regge trajectories, giving an asymptotic representation appropriate for the high-energy part of fixed-$t$ dispersion relations~\cite{Collins:1977jy,Gribov:2009zz}. A typical Regge contribution has the schematic form
\begin{equation}
\mathcal{A}_{R}(s,t) =
\eta_R(t)\,\beta_R(t) \left(\frac{s}{s_0}\right)^{\alpha_R(t)},
\label{eq:regge_schematic}
\end{equation}
where $\alpha_R(t)$ is the Regge trajectory, $\beta_R(t)$ is the residue, $\eta_R(t)$ is the corresponding signature factor, and $s_0$ is a fixed hadron scale, conventionally of order $1~\mathrm{GeV}^2$. The detailed $\pi\pi$ amplitudes are given in Appendix~\ref{app:regge}. Regge theory supplies the full asymptotic fixed-$t$ amplitude, not a finite set of high-energy partial waves. We employ the Regge input  used in the well-known $\pi\pi$ dispersive analysis of Garc\'ia-Mart\'in et al.~\cite{Pelaez:2003ky,Yndurain:2007qm,GarciaMartin:2011jx,GarciaMartin:2011cn,Caprini:2011ky}. The resulting contribution is projected into the Roy driving terms $\mathrm{DT}_\ell^I(s)$. Because the Roy equations are twice subtracted, this high-energy contribution is smooth and strongly suppressed in the low-energy region where the SINN is trained~\cite{Rodas:2023nec}. Given this strong suppression, we keep the Regge input fixed throughout the analysis.

\section{Experimental data}
\label{sec:data}

The global $\pi\pi$ data set used in this work contains extractions of phase shifts and inelasticities at fixed energies for a finite set of partial waves. We include seven waves, $S_0$, $P_1$, $S_2$, $D_0$, $D_2$, $F_1$, and $G_2$, and use phase-shift data for all of them. Inelasticity data are used for $S_0$, $P_1$, $S_2$, $D_0$, and $F_1$, while $D_2$ and $G_2$ are treated as elastic throughout the fitted region due to the lack of reliable inelasticity data.

Most of the historical database comes from experiments performed in the 1970s~\cite{Cohen:1973yx,Durusoy:1973aj,Hyams:1973zf,Losty:1973et,Protopopescu:1973sh,Grayer:1974cr,Estabrooks:1974vu,Hyams:1975mc,Rosselet:1976pu,Wetzel:1976gw,Hoogland:1977kt,Polychronakos:1978ur,Martin:1979gm,Cohen:1980cq,Etkin:1981sg}, with later inputs complementing specific waves and energy ranges~\cite{Kaminski:1996da,Alde:1998mc,BNL-E865:2001wfj,Pislak:2003sv,Batley:2010zza}. As mesons are unstable, it is not possible to produce free-pion target data for direct measurements of the $\pi\pi$ amplitude. Most data for phase shifts and inelasticities are extracted from reactions such as $\pi N\to\pi\pi N^\prime$, where the production mechanism and the extrapolation of the exchanged pion must be modeled before an amplitude can be extracted. This procedure carries systematic uncertainties that are difficult to quantify, and different analyses of related data can lead to incompatible solutions~\cite{Kaminski:2002pe,Kaminski:2006yv,GarciaMartin:2011jx,Ochs:2013gi,Perez:2015pea}. Furthermore, different partial waves are extracted from the same data set and are therefore correlated, but this information is unpublished. Modern low-energy inputs from $K$ decays are cleaner~\cite{BNL-E865:2001wfj,Pislak:2003sv,Batley:2010zza}, but they only cover the very low-energy region.

Several data sets also contain discrete ambiguities~\cite{JointPhysicsAnalysisCenter:2023gku,Fornes:2025ldd,Barrelet:1971pw,Chung:1997qd}. The CERN-Munich and related analyses quote multiple extractions for the same underlying measurements. If mutually incompatible experiments are included on equal footing, the fit quality degrades severely, increasing variance and potentially hiding systematic shifts in derived quantities. In fact, this same problem appears in other scattering reactions, like $\pi K$~\cite{Pelaez:2020gnd}. Traditionally, data has been selected heuristically to produce a subset of the available data, but this is a subjective process that can introduce biases that are difficult to quantify. Finally, we include an independent measurement based on pionium production from the DIRAC experiment, which provides additional constraints on the $S$-wave scattering-lengths at threshold~\cite{Adeva:2011tc}. For consistency with other dispersive works, we do not include data from lattice QCD at the physical point~\cite{RBC:2021acc,RBC:2023xqv}.

We return to the data problem in \cref{sec:selection}, where the SINN-based spectral response clustering methodology is presented.

\section{SINN framework}
\label{sec:method}

We now present the S-matrix informed neural network architecture: a flexible representation of the partial waves on the real axis, constrained by the selected data, threshold behavior, unitarity, and by dispersion relations.

\subsection{Architecture and hybridization}
\label{sec:architecture}

\begin{figure*}[t]
    \centering
    \includegraphics[width=\linewidth]{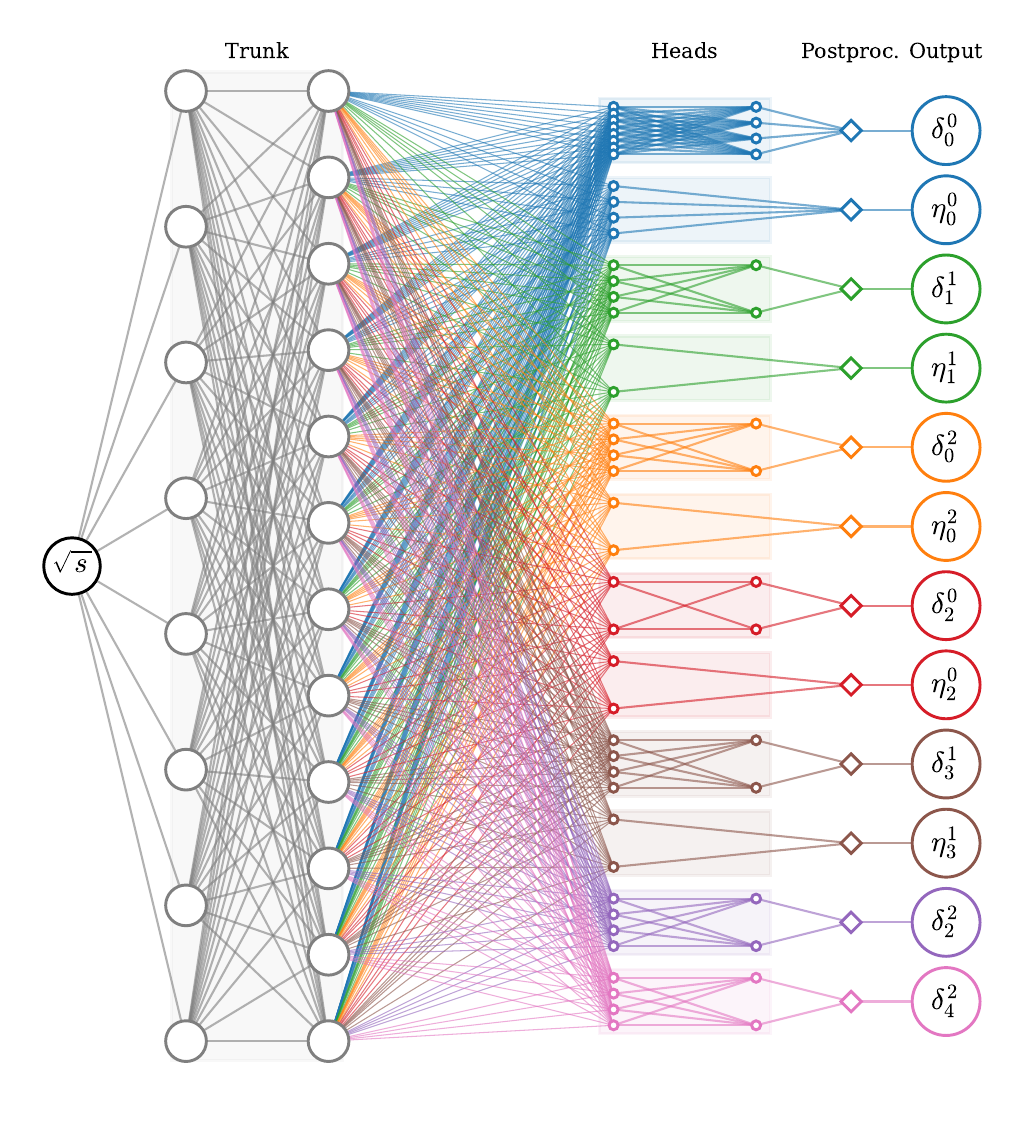}\\
    \caption{SINN architecture employed in our analysis. Each bubble in the Trunk and Heads parts corresponds to a neuron. Postprocessing bubbles correspond to the exact analytic maps of \cref{sec:architecture}: threshold factors and the inelasticity map. Color scheme corresponds to the one used for the observables in subsequent figures.}
    \label{fig:architecture}
\end{figure*}
The input to the SINN is a single number, the energy $\sqrt{s}$ at which one wishes to compute the phase shifts and inelasticities of the scattering amplitude. The network architecture consists of a shared `trunk' neural network connected to multiple `head' neural networks which produce individualized outputs for each observable, followed by a postprocessing layer which directly enforces a selection of hard physical constraints on the output. The final SINN architecture used for $\pi\pi$ analysis is shown in \cref{fig:architecture}, and we refer to it as our `production' SINN.

The shared trunk, split head architecture is motivated primarily by crossing symmetry: The partial waves are not independent objects but projections of a common scattering amplitude, so their neural representation should share information before separating into partial-wave-specific outputs. The shared trunk supplies an inductive bias for correlated partial waves by sharing a real-axis representation before the channel-specific outputs separate. The specialized heads retain the distinct threshold behavior, inelastic openings, and experimental data coverage of each wave. This is a multi-task architecture with hard parameter sharing~\cite{Caruana1997}, and the trunk includes residual connections. The activation functions are chosen smooth, which follows from analyticity and causality. Piecewise-linear activations such as $\mathrm{ReLU}$ interfere with the smoothness of the partial waves spoiling the fulfillment of the Roy equations.

The final postprocessing layer is where hard physics constraints enter the network output. It is a channel-dependent final activation layer that enforces selected constraints exactly. Enforcing these constraints at the output reduces the search space and improves training stability compared with imposing constraints only through the loss. Physics enters both through this postprocessing layer and through specialized loss terms. We refer to the head outputs before postprocessing as `raw' values. These postprocessing constraints are satisfied algebraically. in contrast, analyticity and crossing symmetry are imposed numerically through the Roy-equation loss. The specifics of the postprocessing are as follows.

Each raw phase-shift is multiplied by a threshold factor, which enforces the required behavior at low energies, 
\begin{align}
\delta_\ell^I(s) = \frac{2q^{2\ell+1}}{\sqrt{s}}~\tilde{\delta}_\ell^I(s),
\label{eq:threshold}
\end{align}
where $\tilde{\delta_\ell^I}$ is the raw phase shift. 
Inelasticities are mapped as,
\begin{align}
    \eta_\ell^I(s) = \exp\!\left[-\tilde{\eta}^2\, q_c(s)^{(2\ell_c+1)}\right],
\label{eq:inelasticity}
\end{align}
where $\tilde{\eta}$ is the raw network output, $c$ labels the inelastic channel assigned to the partial wave, and
\begin{align}
q_c(s)&=\frac{1}{2\sqrt{s}}
\sqrt{\big[s-(m_{1,c}+m_{2,c})^2\big]\big[s-(m_{1,c}-m_{2,c})^2\big]}\, \nonumber \\
&\times \Theta\!\left(s-(m_{1,c}+m_{2,c})^2\right).
\label{eq:inelasticity_momentum}
\end{align}
$m_{1,c}$ and $m_{2,c}$ refer to the masses of particles produced by pion scattering in channel $c$. The Heaviside step function $\Theta$ makes $q_c(s)=0$ below the inelastic threshold \mbox{$s_c=(m_{1,c}+m_{2,c})^2$}. The masses defining each effective inelastic channel and the effective momentum $\ell_c$ are summarized in \cref{tab:inel_channels}. The $S_0$ and $D_0$ waves open through $K\bar K$, $P_1$ and $F_1$ through $\pi\omega$, and the $S_2$ wave uses an effective two-particle threshold with $m_{\mathrm{eff}}=0.55~\mathrm{GeV}$. The $D_2$ and $G_2$ waves have no inelasticity output in the production fit and are kept elastic throughout the fitted region. This form guarantees $0 \le \eta_\ell^I \le 1$ by construction, and automatically enforces elastic unitarity below each channel's inelastic threshold~\cite{Watson:1954uc,GarciaMartin:2011cn}.

\begin{table}[!htb]
\caption{Inelastic-channel assignments used in the postprocessing map of \cref{eq:inelasticity}. The $S_2$ threshold uses an effective mass $m_{\text{eff}} = 0.55~\mathrm{GeV}$. The $D_2$ and $G_2$ waves are treated as elastic.}
\label{tab:inel_channels}
\centering
\begin{tabular}{l c c c}
\toprule
Partial wave & $c$ & $\sqrt{s_c}\,(\mathrm{GeV})$ & $\ell_c$ \\
\midrule
$S_0$ & $K\bar{K}$  & $0.9914$ & $0$ \\
$P_1$ & $\pi\omega$ & $0.9222$ & $1$ \\
$S_2$ & effective   & $1.1000$ & $1$ \\
$D_0$ & $K\bar{K}$  & $0.9914$ & $2$ \\
$F_1$ & $\pi\omega$ & $0.9222$ & $2$ \\
\bottomrule
\end{tabular}
\end{table}

The full production SINN architecture is provided in Appendix~\ref{app:hyperparameters}. 

\subsection{Loss design}
\label{sec:loss}

The amplitudes must describe the data while simultaneously satisfying the Roy equations and the low-energy consistency conditions at threshold. This makes the training a multi-objective optimization problem. Network parameters are determined by minimizing the loss, decomposed as sub-loss terms multiplied by curriculum weights, 
\begin{align}
\mathcal{L}_{\text{total}} = \lambda_{\text{data}}\mathcal{L}_{\text{data}} + \lambda_{\text{Roy}} \mathcal{L}_{\text{Roy}} +\lambda_{\mathrm{SL}}\mathcal{L}_{\mathrm{SL}}+\lambda_{\text{con}} \mathcal{L}_{\text{con}}.
\label{eq:total-loss}
\end{align}

The data term is
\begin{align}
    \mathcal{L}_{\text{data}} = \frac{1}{N_{\text{tot}}} \sum_{\alpha} \sum_{i=1}^{N_\alpha} \left[(1-w_\chi)\, r_{\alpha,i}^2 + w_\chi \left(\frac{r_{\alpha,i}}{\sigma_{\alpha,i}}\right)^2\right],
\label{eq:chi2}
\end{align}
where $r_{\alpha,i} = y_{\alpha,i}^{\mathrm{SINN}} - y_{\alpha,i}^{\text{exp}}$, $\alpha$ runs over observable types, and $N_{\text{tot}} = \sum_\alpha N_\alpha$. The parameter $w_\chi$ interpolates between unnormalized least squares ($w_\chi = 0$) and a standard $\chi^2$ ($w_\chi = 1$).

The Roy equation penalty is
\begin{align}
    \mathcal{L}_{\text{Roy}} = \sum_{\ell_I \in \{S_0,P_1,S_2\}} \frac{1}{N_s}\sum_{i=1}^{N_s} \left| \Delta^I_{\ell}(s_i) \right|,
\label{eq:roy-loss}
\end{align}
where $N_s=75$ is the number of points in the fixed energy mesh $\{s_i\}$ at which the Roy residuals, $\Delta^I_\ell(s)$, are computed. These residuals are simply computed by recasting each equation in \cref{eq:roy} as a difference which equals zero when the Roy equations are perfectly satisfied. The residual is evaluated for the three low-energy dominant waves $S_0$, $P_1$, and $S_2$. Higher waves are not assigned Roy residuals; they enter only as input to the dispersive integrals of the three constrained waves.

The final two loss terms both relate to the behavior of the SINN at the two pion threshold. The subtraction constants entering the Roy equations must coincide with the scattering lengths defined by the threshold limit of the same partial waves, see \eqref{eq:ST}, as required for the dispersive representation to be self-consistent at threshold. We introduce additional auxiliary learned subtraction constants $a_0^I$, and compute the output scattering lengths $a_0^{I,\text{SINN}}$ from the SINN amplitude separately. A scattering length consistency loss is imposed,
\begin{align}
    \mathcal{L}_{\text{con}} = \left|a_0^0 - a_0^{0,\mathrm{SINN}}\right| + \left|a_0^2 - a_0^{2,\mathrm{SINN}}\right|,
\label{eq:sl-loss}
\end{align}
with \mbox{$a_0^{I,\mathrm{SINN}}=\lim_{s\to4m_\pi^2}\mathrm{Re}\,t_0^I(s)$}. This term penalizes disagreement between the auxiliary subtraction constants and the low-energy threshold behavior learned by the network. This term was introduced to help stabilize training; the scattering lengths are quantities derived from the network weights, and there is no simple way to initialize them without the inclusion of extra parameters. Poor initializations can cause $\mathcal{L}_{\rm Roy}$ to be dominated by the subtraction terms in the early stages of training.

The DIRAC pionium measurement provides an additional soft constraint on the $S$-wave scattering-length difference at threshold $D_{\text{DIRAC}}\equiv|a_0^0-a_0^2|$~\cite{Adeva:2011tc}, 
\begin{align}    D_{\text{DIRAC}}=\left(\left.0.2533^{+0.0080}_{-0.0078}\right|_\mathrm{stat}
\left.{}^{+0.0078}_{-0.0073}\right|_\mathrm{syst}\right) m_\pi^{-1}~\nonumber.
\end{align}

We penalize disagreement between the SINN-derived combination $\lvert a_0^{0,\mathrm{SINN}}-a_0^{2,\mathrm{SINN}}\rvert$ and this central value via the asymmetric $\chi^2$-style loss 
\begin{align} 
\mathcal{L}_{\mathrm{SL}} = \left( \frac{ \lvert a_0^{0,\mathrm{SINN}}-a_0^{2,\mathrm{SINN}}\rvert - D_{\text{DIRAC}} }{ \sigma_{\pm} } \right)^2, \label{eq:dirac-loss} 
\end{align} 

weighted by $\lambda_{\mathrm{SL}}$ in the total objective, where the combined statistical and systematic uncertainty $\sigma_+$ ($\sigma_-$) is used when the residual is positive (negative). Hence, the scattering lengths are constrained both by agreement between the dispersive and neural representations of the amplitude, and also by the DIRAC measurement of the magnitude of the difference in the lengths. The coupling $\lambda_{\mathrm{SL}}$ is held fixed at $1/N_{\rm{data}}$ so that it is treated as a single data point, identically to the rest of the experimental data.

\subsection{Curriculum training}
\label{sec:curriculum}

\begin{table}[!htb]
\caption{Curriculum weight schedule. In the final phase, the auxiliary trainable subtraction constants are replaced by network-derived values and $\mathcal{L}_{\mathrm{con}}$ is dropped. $\lambda_{\mathrm{SL}}$ is not listed because it is held fixed at $1/N_\mathrm{data}$ in every phase.}
\label{tab:curriculum}
\centering
\begin{tabular}{c c c c c c}
\toprule
Phase &Epochs & $\lambda_{\mathrm{data}}$ & $\lambda_{\mathrm{Roy}}$ & $\lambda_{\mathrm{con}}$ & $w_\chi$ \\
\midrule
I&0--400 & 1 & 0 & 0 & 0.25 \\
II&400--800 & 1 & 0.01 & 0.01 & 0.5 \\
III&800--1000 & 1 & 0.5 & 0.1 & 0.75 \\
IV&1000--1200 & 1 & 1 & 1 & 1 \\
V&1200--1500 & 1 & 1 & 10 & 1 \\
VI&1500--2000 & 1 & 2 & --- & 1 \\
\bottomrule
\end{tabular}
\end{table}

In total, the training minimizes $17$ competing objective functions, composed of the $12$ data terms, $3$ Roy terms, and $2$ threshold scattering-length objectives. The total loss is non-convex, and these objectives cannot all be turned on from the start of the learning process. The Roy equations couple all partial waves through dispersive integrals; if $\mathcal{L}_\mathrm{Roy}$ is active before the network has learned the rough shape of each wave from data, the integral equations dominate the gradient. The optimizer may then converge to a nearly trivial solution that satisfies Roy equations but misses the experimental data entirely. Similarly, quoted experimental errors vary significantly across channels in the global data set, which can cause some partial waves to be completely ignored during training. We use curriculum-based learning combined with multi-objective loss aggregation to avoid these failure modes. We first learn a data-driven amplitude, and gradually tighten the physics constraints while juggling the competing objectives.

The weight curriculum (\cref{tab:curriculum}) contains six phases. The first phase uses only the scattering length loss combined with data loss at weight \mbox{$w_\chi=0.25$} rather than a full $\chi^2$. This allows the fit to learn the shape of every observable before the smallest experimental errors dominate the optimization.

As training proceeds, $w_\chi$ is increased to 1, transforming the data term into the standard error-weighted $\chi^2$. At the same time, the Roy and scattering-length consistency penalties are switched on gradually. The auxiliary Roy subtraction constants $a_0^I$ first stabilize the dispersive equations within the magnitude set by the DIRAC measurement, while $\mathcal{L}_{\mathrm{con}}$ forces them to approach the network-derived threshold values $a_0^{I,\mathrm{SINN}}$. In the final phase, this loss term is fixed to be exactly zero: the auxiliary subtraction constants are removed, and the Roy equations use $a_0^{I,\mathrm{SINN}}$ directly.

\begin{figure*}[t]
\centering
\includegraphics[width=\linewidth]{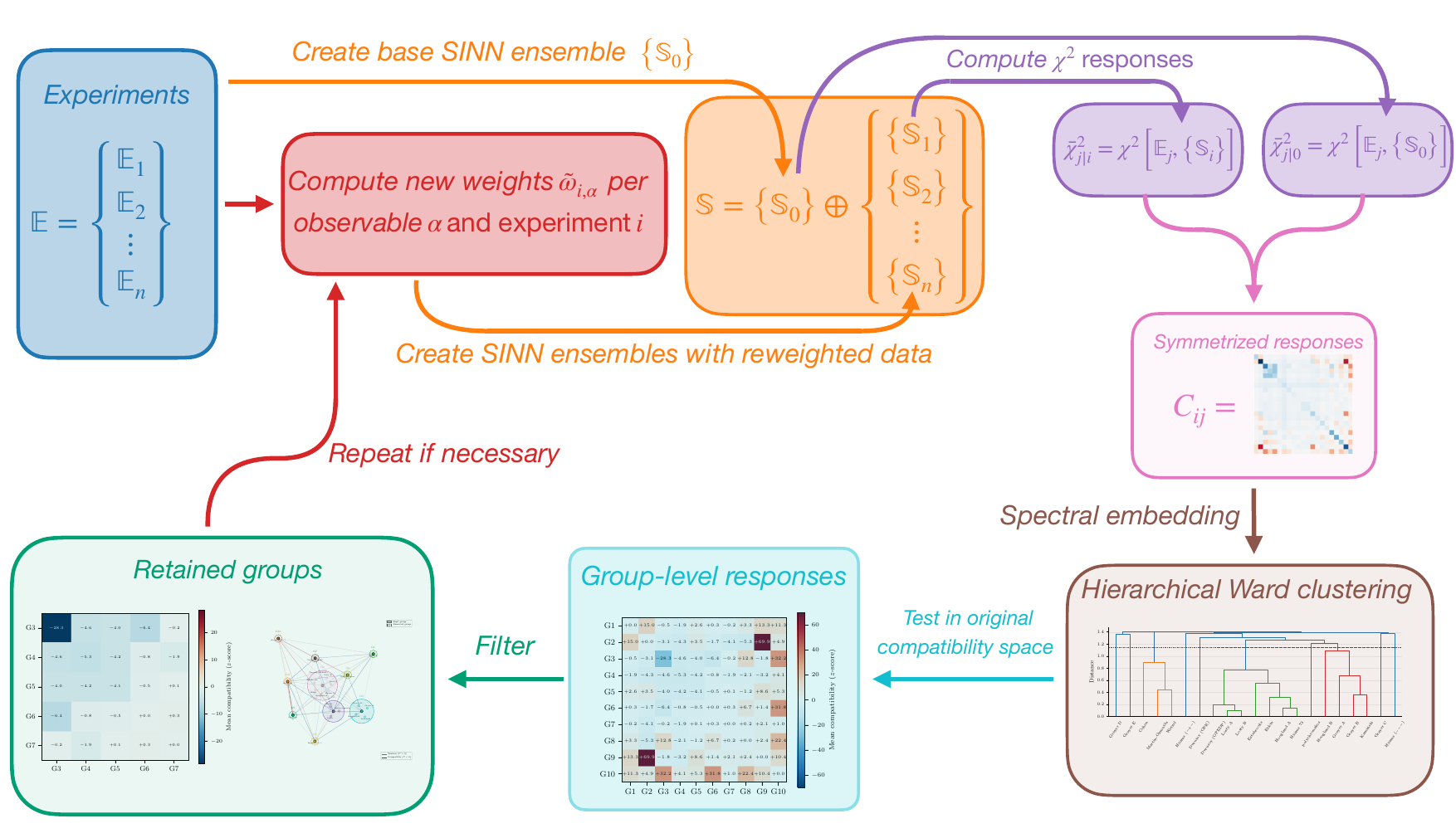}
\caption{Data selection workflow via spectral response clustering. See text for details. $\mathbb{E}$ and $\mathbb{S}$ stand for the sets of all  experiments and SINNs, respectively.}
\label{fig:dataselection}
\end{figure*}

At each optimization step, fine-grained UPGrad combines the active objective gradients separately rather than differentiating only the weighted scalar loss~\cite{quinton2025jacobiandescentmultiobjectiveoptimization}. This projects conflicting objective gradients toward a common descent direction; see Ref.~\cite{lin2025multiobjective} for a survey of gradient-based multi-objective methods. The network parameters are obtained with AdamW optimization~\cite{Kingma2015,Loshchilov2019}, Xavier weight initialization~\cite{Glorot2010}, gradient clipping, weight decay, and patience-based phase-transition and termination criteria. Early stopping with a patience of 50 epochs is active in all phases. In earlier phases, triggering the criterion advances the curriculum to the next phase. Overall, the curriculum improves optimization stability and reduces the number of epochs required for convergence. UPGrad supplies the parameter-update direction from the separately aggregated active objectives. Each epoch is also evaluated with the displayed final-stage scalar loss, using the final curriculum weights and network-derived scattering lengths; the converged model is the epoch with the lowest value of that loss.

The remaining numerical choices are collected in Appendix~\ref{app:hyperparameters}.

\subsection{Ensemble uncertainty quantification}
\label{sec:ensemble}

Uncertainty quantification uses the Gaussian replica method~\cite{Ball:2008by,Ball:2009qv,NNPDF:2017mvq,JPAC:2021rxu}. This method combines a parametric bootstrap of the data with independently trained networks, and is closely related to deep-ensemble uncertainty estimation~\cite{Lakshminarayanan2017}. For each replica, every data point $y_i^{\text{exp}}$ is shifted by \mbox{$\epsilon_i \sim \mathcal{N}(0, \sigma_i^2)$} and a new network is trained on the shifted data. Inelasticity values are clipped to $[0,1]$ after the shift, to avoid providing unphysical resampled data. Each replica is trained from five independent random weight initializations, and the initialization with the lowest final-stage total loss for that replica is kept. The DIRAC data are resampled from the split-normal distribution defined by its asymmetric quoted uncertainties in replica generation.

Every retained replica uses the same hard threshold and unitarity maps, is evaluated with the same final-stage objective, and passes the common total-loss cut. The resulting bands are fully constrained: each interval is an empirical distribution over constrained amplitudes, with every member a separate solution of the same constrained problem. The production intervals directly contain experimental Gaussian replicas, split-normal DIRAC replicas, and fit-initialization variation. The loss distributions over the retained ensemble, the training evolution of a representative replica, and the trained weight distribution are collected in Appendix~\ref{app:supp_validation}.

\section{Data selection}
\label{sec:selection}

Having described the full experimental data set in \cref{sec:data}, we now present the reproducible selection procedure used to generate the production ensemble. We refer to our procedure as \textit{spectral response clustering}, summarized in~ \cref{fig:dataselection}. Its central feature is that constrained SINN ensembles measure the relative compatibility between experiments. In ML, retraining-based influence analysis estimates the effect of individual training data on model behavior~\cite{KohLiang2017}, and spectral clustering of influence patterns obtained from ensembles of retrained models reveals coherent subpopulations of data~\cite{Ilyas2022}. In global parton distribution functions fits, Hessian-based diagnostics quantify pairwise tensions between experiments and embed data sensitivities for visualization~\cite{Wang:2018heo,Jing:2023isu}, complementing the weighted-fit selection of Ref.~\cite{NNPDF:2021njg}. Spectral response clustering combines these ideas: responses are measured at the experiment level by deliberate upweighting inside a physics-constrained model class, and the clustered response matrix drives an automated selection decision.

\subsection{Spectral response clustering}
\label{subsec:selection_strat}

\begin{figure}[!ht]
\centering
\includegraphics[width=\linewidth]{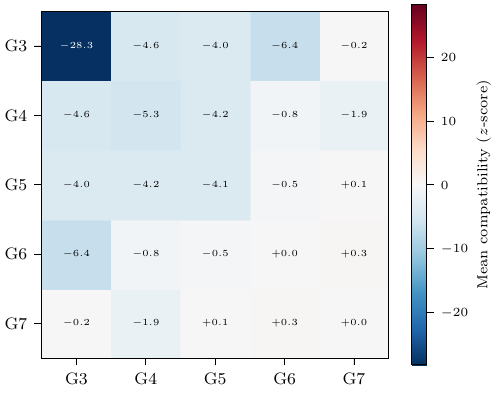}
\caption{Group-level response after selection. The maximum surviving between-group response is $+0.3$.}
\label{fig:group_compat}
\end{figure}

For each target experiment $\mathbb{E}_i$, we construct an ensemble of SINNs in which $\mathbb{E}_i$ is upweighted relative to the rest of the data in every observable for which it quotes data. For each observable $\alpha$ we define
\begin{align}
w_{i,\alpha} = \frac{N_{\text{other},\alpha}}{N_{i,\alpha}},
\end{align}
where $N_{i,\alpha}$ is the number of data points contributed by $\mathbb{E}_i$ to observable $\alpha$ and $N_{\text{other},\alpha}$ is the number of points from all other experiments that contribute to that observable. All other experiments start with unit weight. This gives the target experiment a weight comparable to the combined weights of the remaining data for each individual observable which it contributes data to. For observables where $\mathbb{E}_i$ is the sole experiment, unit weights are retained. The resulting selection weights are globally renormalized as
\begin{align}
\tilde{w}_{j,\alpha} = \eta \cdot w_{j,\alpha}, \quad
\eta = \frac{\sum_{j,\alpha} N_{j,\alpha}}{\sum_{j,\alpha} w_{j,\alpha} N_{j,\alpha}},
\end{align}
so that the total effective number of data points is preserved across the training set, and so that we do not spoil satisfaction of the physics constraints throughout our curriculum-based training. All ensembles are trained with the physics constraints fully on.

For each evaluated experiment $\mathbb{E}_j$, we measure the fit quality of each retained model $r$ in the ensemble $\{\mathbb{S}_i\}$ trained with target experiment $\mathbb{E}_i$ upweighted using the corresponding uniform-weight $\chi^2$,
\begin{align}
\chi^2_{j|i ,r} = \frac{1}{N_j}\sum_{k \in \mathbb{E}_j}
\frac{\big(y_{k,i,r}^{\mathrm{SINN}}-y_k^{\text{exp}}\big)^2}{\sigma_k^2}.
\end{align}
The baseline ensemble, labeled $i = 0$ ($\{\mathbb{S}_0\}$), is trained with the same architecture and curriculum but without experiment-specific upweighting. Let $\bar{\chi}^2_{j|i},~ s_{j|i},$ and $n_i$ denote the sample mean, sample standard deviation, and retained model count for $\chi^2_{j|i,r}$. The directed response is,
\begin{align}
z_{ij} =
\frac{\bar \chi^2_{j|i} - \bar\chi^2_{j|0}}
{\sqrt{s_{j|0}^2/n_{0} + s_{j|i}^2/n_{i}}}~.
\end{align}
Because upweighting $\mathbb{E}_j$ and evaluating $\mathbb{E}_i$ need not give the transposed response, we symmetrize the statistic as
\begin{align}
C_{ij}=\frac{z_{ij}+z_{ji}}{2},
\end{align} 
to produce the {\it symmetrized  response} between experiment $\mathbb{E}_i$ and $\mathbb{E}_j$. Positive values indicate tension, because increasing the weight of one experiment increases the $\chi^2$ of the other, while negative values indicate compatibility. Each response can be thought of as a matrix element of the \textit{response matrix}~$C$.

\begin{figure*}[tb]
\centering
\includegraphics[width=0.85\linewidth]{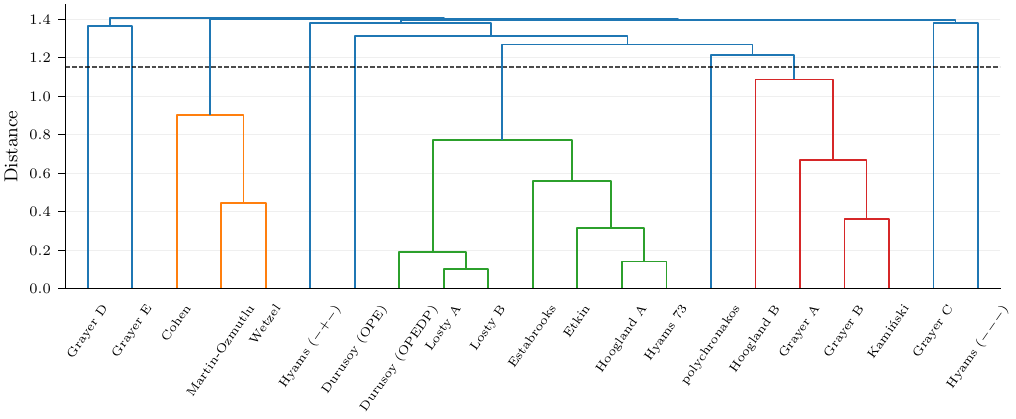}
\caption{Ward-clustering dendrogram for the experiment response matrix. The cut at the largest merge-height gap produces ten groups.}
\label{fig:dendrogram}
\end{figure*}
\begin{table*}[tb]
\caption{Experimental data used in the final fit. The ``Excluded'' column lists experiments removed by the selection procedure.}
\label{tab:data-inventory}
\centering
\resizebox{\textwidth}{!}{%
\begin{tabular}{l l l}
\toprule
Observable & Included experiments & Excluded experiments \\
\midrule
$|a_0^0-a_0^2|$ & DIRAC &  \\
$S_0$ phase & Grayer (A, B), Kami\'{n}ski, $K$ decay, ${\rm NA}_{48}$ & Grayer~(C, D, E), Hyams ($-$$-$$-$, $-$$+$$-$), Protopopescu (VI, XIII) \\
$S_0$ inelasticity & Cohen, Etkin, Grayer~B, Kami\'{n}ski, Martin-Ozmutlu, Polychronakos, Wetzel & Hyams ($-$$-$$-$, $-$$+$$-$), Protopopescu (VI, XIII) \\
$P_1$ phase & Estabrooks, Hyams~73 & Hyams ($-$$-$$-$, $-$$+$$-$), Protopopescu (VI, XIII) \\
$P_1$ inelasticity & Hyams~73 & Hyams ($-$$-$$-$, $-$$+$$-$), Protopopescu (VI, XIII) \\
$S_2$ phase & Cohen, Durusoy (OPE, OPEDP), Hoogland (A,B), Losty (A,B) & --- \\
$S_2$ inelasticity & Cohen, Losty~B & --- \\
$D_0$ phase & Hyams~73 & Protopopescu (VI, XIII) \\
$D_0$ inelasticity & Hyams~73 & Protopopescu (VI, XIII) \\
$D_2$ phase & Cohen, Durusoy (OPE, OPEDP), Hoogland (A,B), Losty (A,B) & --- \\
$F_1$ phase & Hyams~73 & Protopopescu (VI, XIII) \\
$F_1$ inelasticity & Hyams~73 & Protopopescu (VI, XIII) \\
$G_2$ phase & Cohen, Durusoy (OPE, OPEDP), Losty~B &  --- \\
\bottomrule
\end{tabular}}
\end{table*}

After the response matrix is constructed, it is embedded into a reduced Euclidean space through a truncated spectral decomposition so that we may identify patterns in the response. That is, we decompose,
\begin{align}
\sum_j C_{ij}\, v_{ja} = \lambda_a\, v_{ia},
\label{eq:response-eig}
\end{align}
and order the eigenvalues by decreasing $|\lambda_a|$. We represent each experiment $\mathbb{E}_i$ by the vector of these leading loadings,
\begin{align}
\mathbf{x}_i = (v_{i1},\ldots,v_{ik}).
\label{eq:response-embedding}
\end{align}
Ward linkage~\cite{Ward01031963,Murtagh2014} is then applied to the set of experiment embeddings $\{\mathbf{x}_i\}$, building a hierarchy of merges which collapse the individual experiments into groups of experiments. Operationally, each experiment is initially considered as a separate cluster. If clusters $A$ and $B$ contain $n_A$ and $n_B$ experiments with embedding centroids $\boldsymbol{\mu}_A$ and $\boldsymbol{\mu}_B$ in the retained $\mathbb{R}^k$ embedding, merging them increases the total within-cluster sum of squares by
\begin{align}
\Delta(A,B) = \frac{n_A n_B}{n_A + n_B}\,
\big\|\boldsymbol{\mu}_A - \boldsymbol{\mu}_B\big\|^2.
\label{eq:ward-distance}
\end{align}
Starting from $N$ singleton clusters, the method repeatedly merges the pair with the smallest $\Delta(A,B)$, building a dendrogram whose merge heights $d_m$ increase monotonically. Denoting the sorted merge heights by \mbox{$d_1 \le d_2 \le \cdots \le d_{N-1}$}, where $N$ is the number of experiments, a cut between $d_m$ and $d_{m+1}$ leaves \mbox{$N-m$} groups. We place the cut at the largest gap between successive heights,
\begin{align}
m^\star = \underset{m}{\arg\max}\,\left(d_{m+1}-d_m\right), \qquad n_{\mathrm{g}} = N - m^\star,
\label{eq:dendrogram-cut}
\end{align}
with $m$ restricted so that \mbox{$2 \le n_{\mathrm{g}} \le N$}, yielding the candidate groups \mbox{$G_1,\ldots,G_{n_{\mathrm{g}}}$}.

The groups are then tested back in the original response space through the averages,
\begin{align}
\bar{C}_{\ell,\ell} &=
\frac{2}{|G_\ell|(|G_\ell|-1)}
\sum_{\substack{i,j \in G_\ell \\ i<j}} C_{ij}, \\
\bar{C}_{\ell,m} &=
\frac{1}{|G_\ell|\,|G_m|}
\sum_{i \in G_\ell}\sum_{j \in G_m} C_{ij},
\end{align}
where $G_\ell$ denotes the set of experiments in group $\ell$. The quantity $\bar{C}_{\ell,\ell}$ is the mean pairwise response inside a group, while $\bar{C}_{\ell,m}$ is the mean pairwise response between groups. Since positive values of $C_{ij}$ indicate tension and negative values indicate compatibility, the same interpretation applies to the group averages. 

\begin{figure*}[t]
\centering
\includegraphics[width=0.9\linewidth]{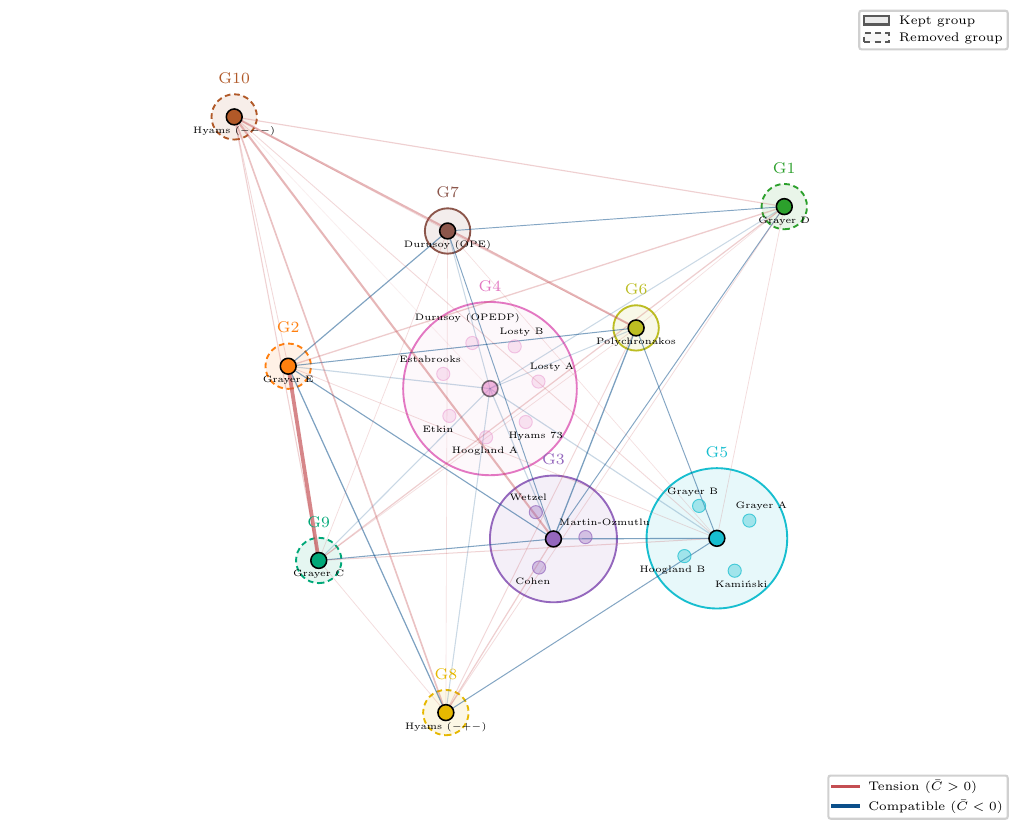}
\caption{Force-directed tension graph of the experiment groups, rendered as a 2D representation of the experiment embeddings. Nodes are individual experiments within their groups (shaded central circles). The layout pushes (in)compatible groups (apart) closer. Red edges connect groups in tension and blue edges connect compatible groups. Retained groups are drawn with solid outlines and filled markers, while rejected groups are drawn with dashed outlines.}
\label{fig:tension_graph}
\end{figure*}

Finally, groups are removed iteratively until no positive surviving between-group average exceeds the cutoff $\bar C_{\rm cut}=1$. The full procedure may then be repeated on the surviving experiments for further reduction in data.

\subsection{Retained experimental input}
\label{subsec:retained}

\begin{figure*}[!ht]
\centering
\includegraphics[width=0.95\linewidth]{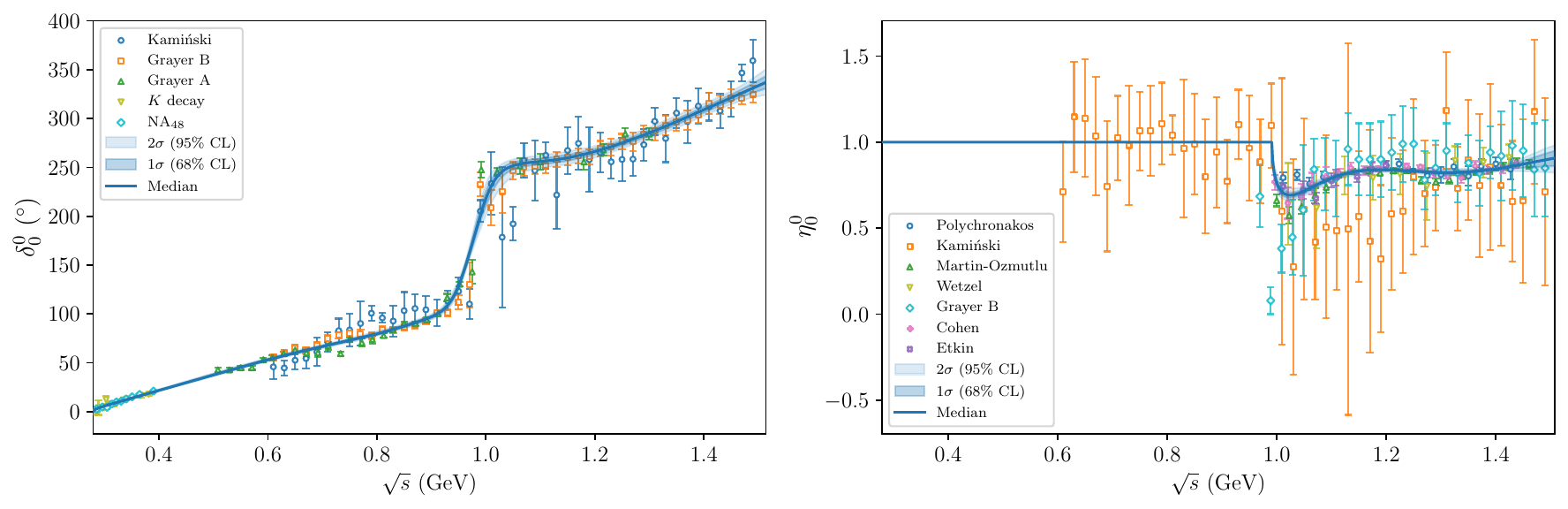}
\includegraphics[width=0.95\linewidth]{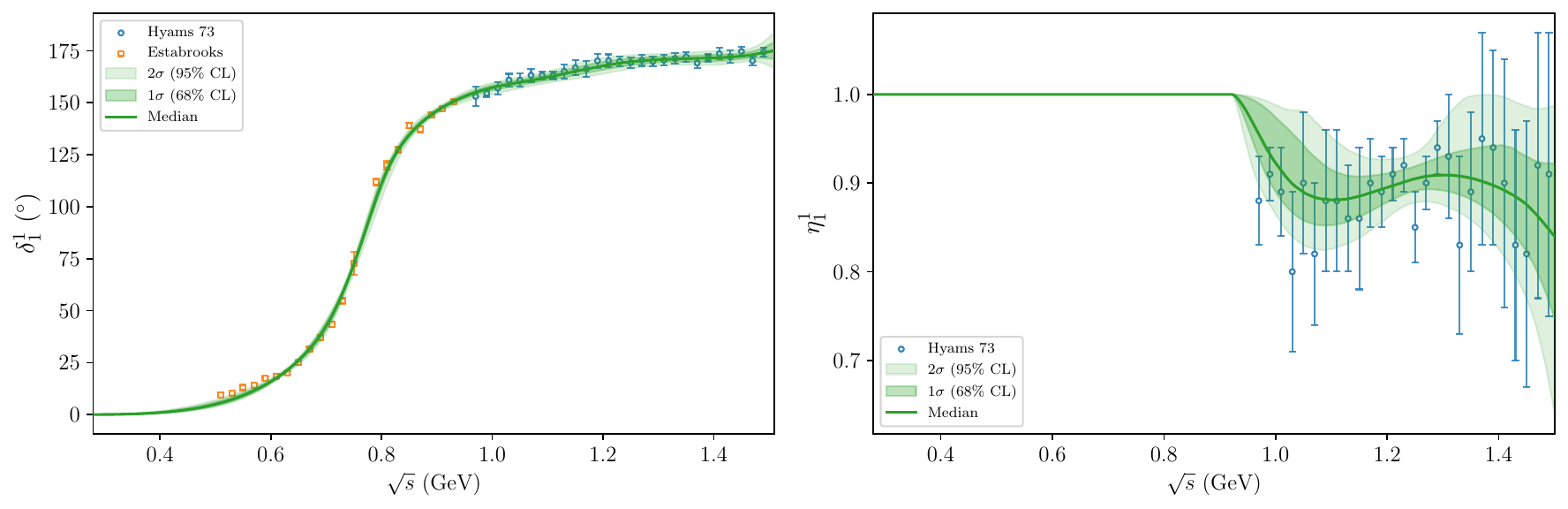}
\includegraphics[width=0.95\linewidth]{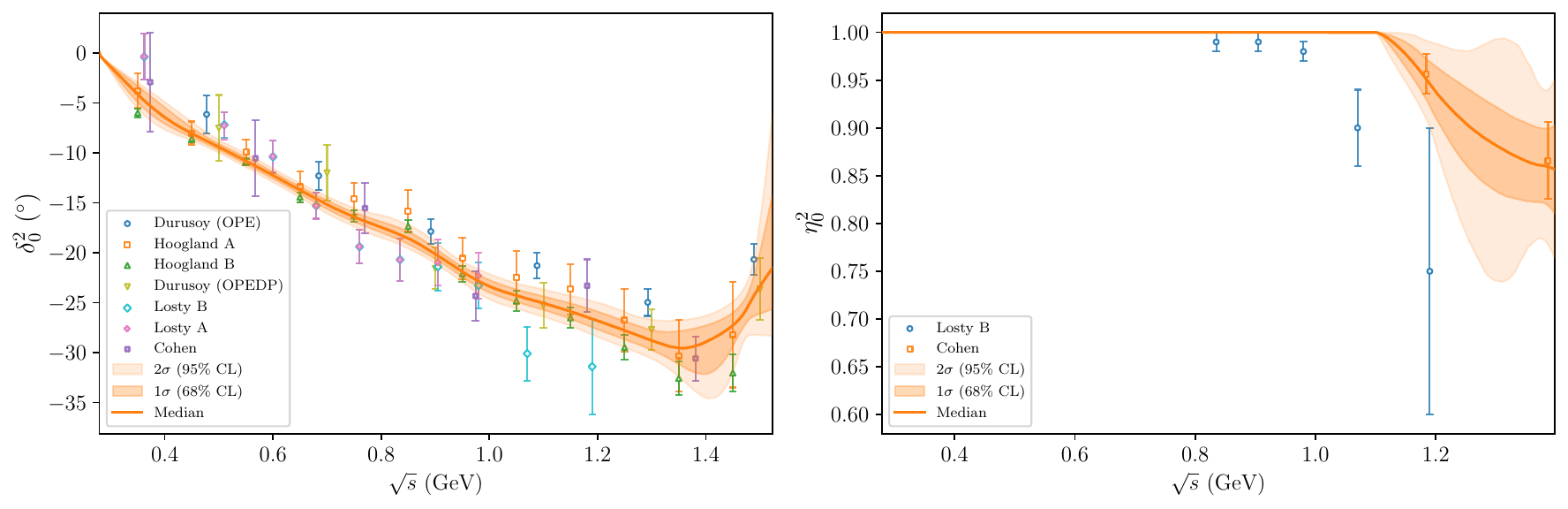}
\caption{$S_0$, $P_1$, and $S_2$ phase shifts and inelasticities. Solid lines are ensemble medians. Dark and light bands are central 68\% and 95\% replica intervals.}
\label{fig:waves_main}
\end{figure*}

Two rounds of spectral response clustering were performed to select the data used to train our production SINN ensemble. The $K$-decay and ${\rm NA}_{48}$ data are retained \textit{a priori} because they supply the highest-quality near-threshold information. The DIRAC data is also retained as it is the only measurement of its kind. The first selection round ran the complete procedure on all other available data. This round immediately removed the Protopopescu (VI, XIII) data sets, which are grossly incompatible with the rest of the data. This can be seen immediately from \cref{fig:s0_data}, as both Protopopescu data sets claim $S_0$ phase shift values far below the entire rest of the data set in the region just above $1~\mathrm{GeV}$. 

The spectral decomposition, Ward clustering, and iterative group-removal steps were repeated on the eligible remaining experiments in a second round of the procedure. The full step-by-step account of this second round is given in Appendix~\ref{app:selection_eigenspace}. The resulting Ward dendrogram is shown in \cref{fig:dendrogram}. Spectral response clustering produced ten groups, and the following five singleton groups were discarded:
\begin{align*}
\mathrm{G1}&:\ \text{Grayer D},\\
\mathrm{G2}&:\ \text{Grayer E},\\
\mathrm{G8}&:\ \text{Hyams }(-\!+\!-),  \\
\mathrm{G9}&:\ \text{Grayer C}, \\
\mathrm{G10}&:\ \text{Hyams }(-\!-\!-).
\end{align*}

The retained data contains $18$ experiments and $621$ measurements in the range \mbox{$0.28 \le \sqrt{s} \le 1.5~\mathrm{GeV}$}. The full data inventory and selection summary is given in~\cref{tab:data-inventory}. Across the two rounds the median total loss of the production architecture falls from $7.2$ in the baseline ensemble to $3.1$ on the retained set. Almost all of this difference is due to the data term, which falls from $6.9$ to $2.9$. The selection procedure could in principle be repeated for a third iteration, though due to the known data quality issues, further iteration would likely serve to artificially lower the statistical uncertainty while increasing selection bias.

During the development of this work, we repeated this procedure with several candidate SINN architectures and produced retained data sets that differed little, if at all, which points to mild dependence on the particular neural representation. providing an empirical check that the selection is not driven by one particular neural representation.

\Cref{fig:tension_graph} gives a global view of the tension structure that drives this procedure in the second data-selection iteration. Each experiment is a node, grouped into its Ward cluster (shaded regions), and the clusters are arranged by a force-directed layout in which mutually compatible groups attract while groups in tension repel. The two-dimensional layout is mainly illustrative, as clustering and rejection are performed in the retained high-dimensional spectral space and original response matrix-space respectively. However, the colored edges make the rejection logic visible: The discarded singleton groups sit on the periphery, whereas the retained groups form an internally blue-linked core.

\section{Results}
\label{sec:results}

\begin{figure*}[t]
\centering
\includegraphics[width=0.95\linewidth]{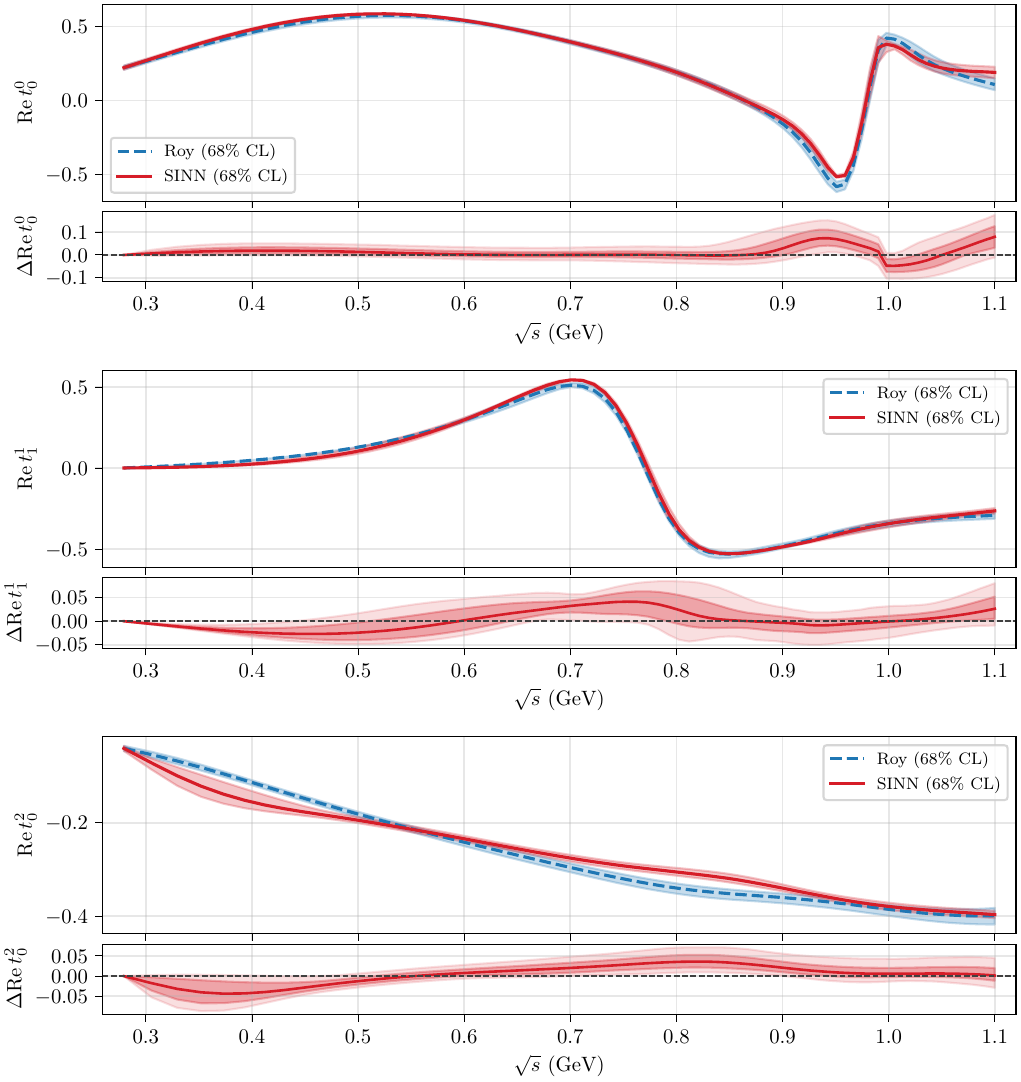}
\caption{SINN $t_\ell^I$ amplitudes (red) compared to Roy-reconstructed $\tilde{t}_\ell^I$ amplitudes (blue, dashed) for $S_0$, $P_1$, and $S_2$, with the residual below each panel. Agreement at the percent level throughout the Roy validity range confirms that crossing symmetry is well satisfied.}
\label{fig:nn_vs_roy}
\end{figure*}

With the compatible experimental input and constrained SINN ensemble established, we now present the production line shapes, their dispersive fulfillment, and the derived scattering lengths extracted from the constrained amplitude.

\subsection{Partial-wave amplitudes}
\label{sec:wave_results}

\Cref{fig:waves_main} shows the $S_0$, $P_1$, and $S_2$ phase shifts and inelasticities. The remaining waves ($D_0$, $D_2$, $F_1$, $G_2$) are shown in Appendix~\ref{app:supp_results}. Solid lines are ensemble medians, and the dark and light shaded bands are central 68\% and 95\% uncertainty bands. Because all waves are retained jointly for each replica, their distributions preserve the cross-wave correlations not present in traditional analyses.

\begin{figure*}[!ht]
\centering
\includegraphics[width=0.48\linewidth]{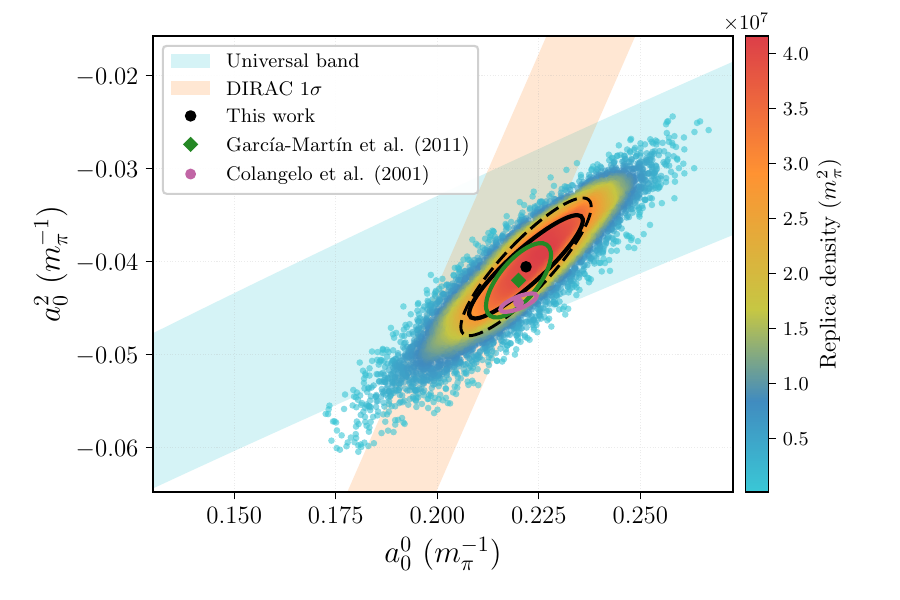} 
\includegraphics[width=0.48\linewidth]{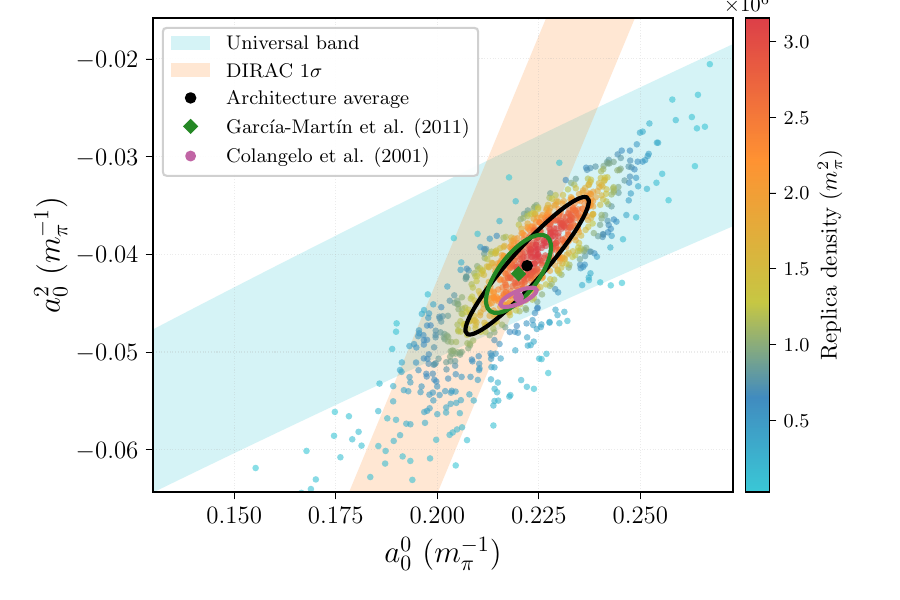}
\caption{(Left panel) Joint distribution of the isospin scattering lengths $a_0^0$ and $a_0^2$ across the $10{,}000$ replicas of the production SINN ensemble. (Right panel) the same distribution for the $1{,}000$-network architecture ensemble of \cref{sec:validation_architecture}, spanning $100$ distinct architectures. In both panels the solid ellipse has half-extents equal to the central $68\%$ interval of that ensemble. In the left panel the dashed ellipse adds the spread of the $100$ per-architecture medians in quadrature, as an approximation of systematic uncertainty in the production interval. Reference determinations from Refs.~\cite{Colangelo:2001df,GarciaMartin:2011cn} are shown for comparison.}
\label{fig:scatt}
\end{figure*}

The $S_0$ phase shift rises through $90^\circ$ near the $\sigma/f_0(500)$ and passes through $180^\circ$ near the $f_0(980)$, while the inelasticity drops sharply at the $K\bar{K}$ threshold. These are all expected features of the $S$ wave. The $P_1$ wave shows the narrow $\rho(770)$. The $S_2$ phase shift is small and negative over the whole range, consistent with the repulsive $I=2$ $\pi\pi$ interaction. The $D_0$ and $F_1$ waves rely on Hyams~73 data alone, and the uncertainty bands are correspondingly wider. Across all waves, the large majority of selected data points lie within the central 68\% uncertainty band. The qualitative features of these bands agree with modern constrained fits to data that satisfy dispersion relations~\cite{Yndurain:2007qm,GarciaMartin:2011jx,Kaminski:2006qe,Kaminski:2011vj,Pelaez:2019eqa,Pelaez:2024uav}.

The $S_0$ inelasticity shows a depletion right above the $K\bar{K}$ threshold, as found in previous dispersive work \cite{GarciaMartin:2011cn,Moussallam:2011zg}. In our analysis, however, the dip is milder. This difference is consistent with two features we find. First, the global selection procedure removes Protopopescu~(VI, XIII), which are among the strongest deep-dip inputs in the historical database. Second, the SINN representation for the phase shift is smooth across the $K\bar{K}$ threshold, in contrast to those analyses. We do not force the model to contain an explicit threshold kink in the $S_0$ phase as the available data is not sufficiently dense in that kinematical region to resolve such a discontinuity in the derivative.

\subsection{Dispersive fulfillment}
\label{sec:dispersive_fulfillment}

Dispersive fulfillment is the central test that the fitted SINN ensemble satisfies the imposed S-matrix constraints. We compare the directly evaluated SINN with the Roy-reconstructed amplitude through \cref{eq:roy}. This is a nonlocal test: the reconstructed $S_0$, $P_1$, and $S_2$ depend not only on the corresponding channel, but on the coupled set of partial waves, the subtraction constants, and the high-energy Regge input as stated before. 
\Cref{fig:nn_vs_roy} confirms this agreement throughout the Roy validity range \mbox{$\sqrt{s} < 1.1~\mathrm{GeV}$} for the $S_0$, $P_1$, and $S_2$ channels, with small average residuals. The $S_0$ and $S_2$ waves share information at threshold by construction, through the subtraction constants. The $P_1$ wave starts at zero because of the barrier factor. The SINN also describes the $S_0$ data near threshold, supporting the extracted scattering lengths. The difference between the SINN amplitude and Roy reconstructed amplitude is consistent with zero across the entire fitted energy range. These comparisons establish numerical fulfillment of the Roy equations for the three constrained waves throughout the domain of validity.

\subsection{Scattering lengths}
\label{sec:scattering_lengths}

The scattering lengths are byproducts of the fit, determined by the SINN amplitude at threshold. They connect directly to the predictions of chiral perturbation theory and, at the same time, provide the subtraction constants for the Roy equations. 

\begin{table}[!ht]
\caption{Scattering lengths in units of $m_\pi^{-1}$. The uncertainty on this work is the central $68\%$ replica interval.}
\label{tab:scattering_lengths}
\centering
\small
\begin{tabular}{c c c c}
\toprule
& This work & Ref.~\cite{Colangelo:2001df} & Ref.~\cite{GarciaMartin:2011cn} \\
\midrule
$a_0^0$ & $0.222^{+0.014}_{-0.014}$ & $0.220 \pm 0.005$ & $0.220 \pm 0.008$ \\[2pt]
$a_0^2$ & $-0.041^{+0.005}_{-0.006}$ & $-0.0444\pm0.0010$ & $-0.042\pm 0.004$ \\
\bottomrule
\end{tabular}
\end{table}

As shown in \cref{tab:scattering_lengths,fig:scatt}, both $a_0^0$ and $a_0^2$ agree within errors with the chiral-Roy determination of Colangelo et al.~\cite{Colangelo:2001df} and the Roy-equation result of \mbox{Garc\'ia-Mart\'in} et al.~\cite{GarciaMartin:2011cn}. Our uncertainties are larger than in traditional dispersive analyses, as we expect to have sampled a broad family of amplitudes compatible with the experimental data and first principles. We do not use a model to constrain the values of the scattering lengths. No SINN which survives the full training curriculum contains parameters associated with the scattering lengths, hence our reported values are constrained predictions. The ensemble also sits inside the so-called ``universal band''~\cite{Ananthanarayan:2000ht}, the region of the scattering-length plane in which solutions of the $\pi\pi$ Roy equations are known to exist, and inside the band that the DIRAC pionium lifetime measurement allows~\cite{Adeva:2011tc}.

Complementary S-matrix-bootstrap studies offer a theory-driven counterpart: they map the space of amplitudes allowed by unitarity, analyticity, and crossing, and now provide dynamical predictions for several QCD observables. Our scattering lengths are compatible with their allowed regions, demonstrating nontrivial agreement between highly dissimilar approaches~\cite{Guerrieri:2018uew,Guerrieri:2024jkn}.

\section{Robustness and validation}
\label{sec:validation}

We next test the robustness of the production ensemble along three complementary axes. First, we test whether the production architecture is representative of the broader family of SINN architectures that can describe the retained data, by comparing against an ensemble of alternative configurations. Second, we perform a closure test, applying the full reconstruction pipeline to synthetic data generated from a known amplitude, to check that the pipeline is unbiased. Third, we retrain the production ensemble with the Roy constraints removed, isolating the role the dispersive constraints play in our analysis. Together, the successful outcome of the three studies supports the interpretation that our SINN ensemble provides a representative sampling of the space of physically allowed amplitudes.

\subsection{Architectural dependence}
\label{sec:validation_architecture}

A neural representation of the S-matrix is only useful if it is flexible enough to describe the data without injecting hidden model dependencies set by the implicit regularization of a particular network. We test for such dependencies with a $500$-trial sweep over $21$ hyperparameter dimensions, covering the trunk geometry, the per-channel head structure, the optimizer settings, the network weight decay, and the Roy weight $\lambda_{\mathrm{Roy}}$. Because $\lambda_{\mathrm{Roy}}$ is itself swept, the trials cannot be compared on their raw objective values; we therefore re-score every trained network on the production final-phase objective before ranking. From the re-scored sweep we retain the $100$ best architectures and train $10$ replicas for each, giving an ensemble of $1{,}000$ networks. Parameter counts across the retained set range from a few hundred to several thousand, against $1{,}156$ for the production network. The construction is detailed in Appendix~\ref{app:sweep_protocol}.

\Cref{fig:scatt} (right) shows the joint distribution of the extracted scattering lengths across the architecture ensemble, and the values are displayed in \cref{tab:robustness}. Every uncertainty quoted in this work is the central $68\%$ interval of the distribution of an observable across the ensemble it is extracted from, taken about the median and quoted asymmetrically; we write $\delta x$ for that interval on an observable $x$. The medians of every single observable are statistically indistinguishable from the production ensemble: the largest shift is $0.42\,\delta x$, and most lie well below that. Both ensembles are built from data replicas, so the architecture-averaged ensemble carries the production statistical spread with architecture variation on top of it. Our central values are insensitive to architecture variation, and we therefore neglect these systematic uncertainties.

\begin{table}[!htb]
\caption{Architecture dependence of the scattering lengths. Values are medians with the central $68\%$ interval taken over every member of each ensemble, the full $10{,}000$-replica production ensemble and the $1{,}000$-network architecture ensemble.}
\label{tab:robustness}
\centering
\small
\begin{tabular}{l c c}
\toprule
Observable & Production & Model average\\
\midrule
$a_0^0$ ($m_\pi^{-1}$)       & $0.222^{+0.014}_{-0.014}$  & $0.222^{+0.013}_{-0.017}$ \\[2pt]
$a_0^2$ ($m_\pi^{-1}$)       & $-0.041^{+0.005}_{-0.006}$ & $-0.041^{+0.005}_{-0.009}$ \\
\bottomrule
\end{tabular}
\end{table}

\subsection{Closure test}
\label{sec:validation_closure}

\begin{table}[t]
\caption{Closure-test results for the $1{,}000$-replica CFD ensemble. Scattering lengths are in units of $m_\pi^{-1}$.}
\label{tab:closure}
\centering
\begin{tabular}{l r r r}
\toprule
Observable & Target & SINN Closure & $z$ \\
\midrule
$a_0^0$                       & $0.2207$  & $0.2201^{+0.0102}_{-0.0104}$ & $-0.06$ \\[2pt]
$a_0^2$                       & $-0.0432$ & $-0.0421^{+0.0028}_{-0.0032}$ & $+0.37$ \\[2pt]
$|a_0^0-a_0^2|$                 & $0.2639$  & $0.2623^{+0.0079}_{-0.0075}$ & $-0.22$ \\[2pt]
\bottomrule
\end{tabular}
\end{table}

\begin{figure*}[t]
\centering
\includegraphics[width=0.92\linewidth]{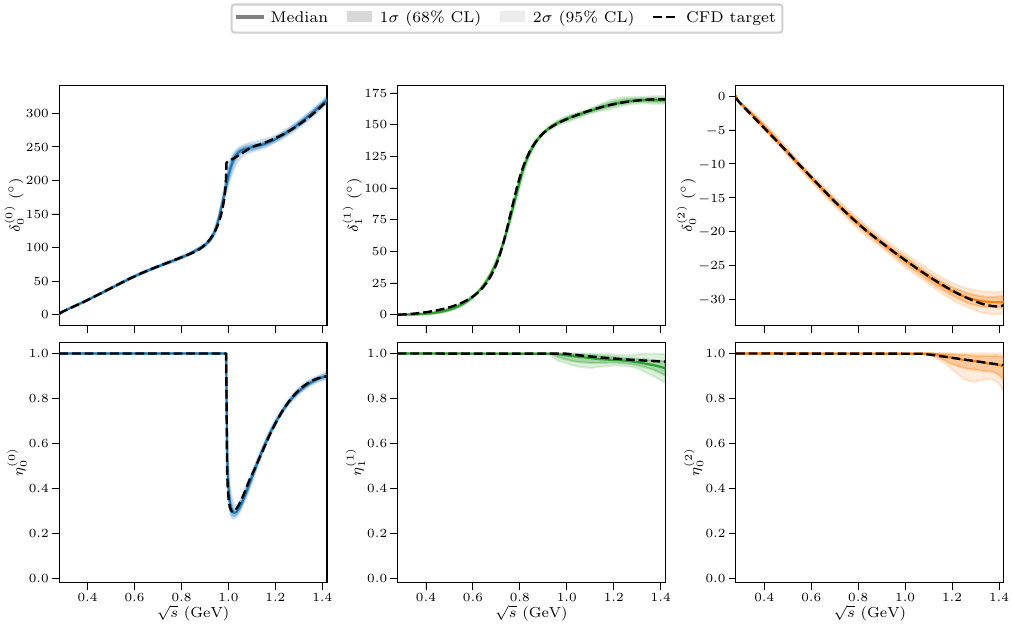}
\caption{Closure test, real-axis reconstruction of the $S_0$, $P_1$, and $S_2$ waves. Solid lines are ensemble medians with $68\%$ and $95\%$ bands; dashed lines are the CFD amplitude used to generate the pseudo-data. Upper row, phase shifts; lower row, inelasticities.}
\label{fig:closure_lineshapes_low}
\end{figure*}

Following the validation strategy of the NNPDF collaboration ~\cite{NNPDF:2021njg,DelDebbio:2021whr}, and simulation-based calibration~\cite{Talts:2018sbc}, we generate pseudo-data from a known parametrization of $\pi\pi$ scattering, the Constrained Fit to Data (CFD) parametrization~\cite{GarciaMartin:2011cn}, at the exact kinematics of the selected data set and with the published uncertainties preserved, then pass them through the full reconstruction pipeline. Because the generating amplitude is known, any displacement of the reconstructed ensemble away from it measures the bias introduced by the SINN representation of the scattering amplitude, and our training strategy. If our production model can accurately reproduce line shapes and extracted observables with data sampled from a fixed parametrization, then we are reconstructing an unbiased $\pi\pi$ amplitude correctly in our production ensemble applied to the selected data.

For an observable with closure-ensemble median $\hat{x}_{\rm closure}$ and target value $x_{\rm target}$, we define the pull,
\begin{align}
z \equiv \frac{\hat{x}_{\rm closure} - x_{\rm{target}}}{\delta x_{\rm closure}},
\label{eq:closure_pull}
\end{align}
so $|z|\le1$ means the ensemble median is compatible with the target value. Results are collected in \cref{tab:closure}. All target values are re-calculated from our implementation of the Roy equations so that no numerical discrepancies may appear between the closure ensembles extraction of an observable and the target value of that observable.

All three scattering-length combinations are recovered near the ensemble centers, with $|z|<0.4$ and every target value inside its asymmetric central $68\%$ interval.

\Cref{fig:closure_lineshapes_low} compares the reconstructed $S_0$, $P_1$, and $S_2$ waves with the CFD amplitude that generated the data. The agreement is excellent across every wave and nearly the whole fitted range. The bands widen only where the pseudo-data inherit sparse or imprecise coverage from the real data set. Appendix~\ref{app:closure} records how the pseudo-data were generated and shows the remaining waves. The one qualitative discrepancy is the $S_0$ phase near the $K\bar K$ threshold, where the CFD phase carries an explicit cusp while the production representation retains the $K\bar K$ inelasticity opening but inserts no phase cusp. Otherwise, the production SINN ensemble fully reproduces the CFD results.

\subsection{Ablation tests}
\label{sec:validation_ablation}

A flexible network fitted to real-axis data will describe that data whether or not the dispersion relations are enforced. To assess the effects of the Roy equations, we retrain a $1{,}000$-replica ablated ensemble with $\lambda_{\mathrm{Roy}},\lambda_\mathrm{con}=0$, holding the architecture, curriculum, and data fixed. \Cref{tab:ablation} collects the comparison between this ensemble and our production ensemble.

\begin{table}[!htb]
\caption{Production versus ablated SINN ensembles, $10{,}000$ and $1{,}000$ replicas. Intervals are central $68\%$. The ablated ensemble carries no total-loss cut.}
\label{tab:ablation}
\centering
\small
\setlength{\tabcolsep}{4pt}
\begin{tabular}{l c c}
\toprule
 & Production & Ablation \\
\midrule
$\mathcal{L}_{\text{data}}$   & $2.93^{+0.21}_{-0.19}$ & $2.93^{+0.23}_{-0.21}$ \\[2pt]
$\mathcal{L}_{\text{Roy}}$ & $0.067^{+0.013}_{-0.011}$ & $0.74^{+0.30}_{-0.41}$ \\
\midrule
$a_0^0$ ($m_\pi^{-1}$) & $0.222^{+0.014}_{-0.014}$ & $0.166^{+0.035}_{-0.027}$ \\[2pt]
$a_0^2$ ($m_\pi^{-1}$) & $-0.041^{+0.005}_{-0.006}$ & $-0.097^{+0.042}_{-0.019}$ \\
\bottomrule
\end{tabular}
\end{table}

The ablated ensemble and production ensemble are indistinguishable based off data satisfaction. Their $\mathcal{L}_{\mathrm{data}}$ medians differ by well under one percent, far inside the spread of either ensemble. The Roy residual is significantly larger on the ablated ensemble, which is expected as that ensemble is Roy-agnostic. The scattering lengths are significantly shifted and exhibit much larger uncertainty bands: $a_0^0$ falls from $0.222$ to $0.166$ and $a_0^2$ from $-0.041$ to $-0.097$, both far outside the production intervals and outside the unviversal band of acceptable scattering lengths. Imposing the Roy equations on a SINN costs nothing in terms of numerical satisfaction of the data, while significantly regularizing the threshold behavior of the amplitude.

\section{Discussion and outlook}
\label{sec:discussion}

We have shown that SINN can accurately describe $\pi\pi$ scattering data, and enables fully constrained data-selection through spectral response clustering. All S-matrix principles are reasonably satisfied through every step of our workflow, and our constrained uncertainty bands carry physical meaning as they estimate the space of physically allowed variations in each line shape and observable. 

Our robustness and validation studies show strong signals that our uncertainties are statistically sound. The closure tests indicates that we are not significantly biasing our results through the specific analysis choices made during data-selection or hyperparameter selection. The only failure of the closure test is the $K\bar K$ cusp in the $S_0$ phase shift, which we deliberately chose not to impose. The ablation study removes the Roy equations entirely from the analysis, and finds the data description untouched. The dedicated sweep over hyperparameters found no shift in the central value of any observable, so we assign no separate uncertainty to the choice of model architecture. 

One could expand the scope of the hyperparameter dependence tests, or attempt a diffusion across architectures and hyperparameters for true model independence. In principle, the selected data set analyzed in our production ensemble implicitly depends on the very same architecture used to select it. Model architecture sweeps and dedicated hyperparameter studies could be performed between each stage of spectral response clustering to exactly quantify any systematics arising between the coupling of the selection model and the selected data set. During testing, several candidate production models were used for spectral response clustering, though the selected data sets across analyses differed very little, if at all. Given the insensitivity of the central values of all extracted observables to model architectures when trained on the final data set, we did not judge that a re-analysis of the selected data sets incorporating these effects would be valuable. Although the number of parameters in a viable SINN architecture is remarkably low by modern machine learning standards, the combinatorics of hyperparameters, architectures, and upweighted ensembles would make such a study more expensive than it is worth. Much of the retained input may carry unknown systematic uncertainties~\cite{Grayer:1974cr,Hyams:1973zf,Hyams:1975mc,Protopopescu:1973sh,Estabrooks:1974vu,Cohen:1980cq,Etkin:1981sg,Alde:1998mc,Kaminski:1996da,Kaminski:2002pe,GarciaMartin:2011jx,Perez:2015pea}, and the kinematic coverage is incomplete. We believe the unknown systematic uncertainties stemming from the poor data quality eclipses notional epistemic uncertainty from model selection during the data selection process. New data from the planned ${\pi}20$ beam line at J-PARC may improve the quality of the amplitude and extracted physics~\cite{JPARC:2026pi20, jparc}

\begin{figure}[!t]
\centering
\fbox{%
\begin{minipage}{0.92\linewidth}
\small
\vspace{2pt}
\textbf{The SINN workflow}
\begin{enumerate}[leftmargin=1.4em, itemsep=3pt, topsep=3pt]
\item \textbf{Split hard and soft constraints.} Separate constraints into those enforceable directly on outputs and those that are soft/nonlocal. \textit{Here: threshold kinematics and unitarity vs. nonlocal dispersion relations.}
\item \textbf{Hybridize neural architecture.} Use parameter sharing as an inductive bias and apply hard constraints through postprocessing. \textit{Here: all outputs are correlated via the architecture; threshold kinematics and unitarity bounds are imposed through postprocessing.}
\item \textbf{Gradually introduce soft constraints.} Train on data first, then introduce soft constraints through a loss curriculum with multi-objective gradient handling. \textit{Here: the six-phase curriculum of \cref{tab:curriculum} with UPGrad.}
\item \textbf{Let the ensemble select the data.} Upweight one experiment at a time, measure the response of every other experiment inside the constrained ensemble, cluster the symmetrized response matrix, and iteratively remove groups in tension. \textit{Here: the two-round spectral response clustering of \cref{sec:selection}.}
\item \textbf{Quantify uncertainty with constrained replicas.} Train an ensemble on resampled data and retain members under a common constrained objective. \textit{Here: $10{,}000$ replicas, including Gaussian experimental and split-normal DIRAC resampling, produce correlated amplitude intervals that can be propagated replica by replica.}
\item \textbf{Estimate systematics.} Repeat previous steps for variations in model architecture and training hyperparameters to estimate bias and control systematic errors. \textit{Here: the architectural dependence, closure, and ablation studies of \cref{sec:validation}}
\end{enumerate}
\vspace{2pt}
\end{minipage}}
\caption{Domain-agnostic summary of the SINN workflow. Each step names the general method, with its realization in the present $\pi\pi$ analysis in italics.}
\label{fig:recipe}
\end{figure}

Several aspects of our analysis transfer directly to other analyses that share similar challenges. First, the combination of the hybridized SINN architecture and training strategy allows the joint encoding of soft and hard physics constraints in flexible neural representations, suitable for other physics-driven inference problems~\cite{Aarts:2025gyp}. Second, the combination of spectral response clustering and replica network ensembles offers a practical method for data selection and uncertainty quantification in constrained systems without sacrificing interpretability of the resulting error bands. Together these aspects form the basis of the SINN workflow, as shown in \cref{fig:recipe}. We anticipate that this workflow may be easily adopted to perform robust analysis of other scattering processes, or even analysis of other systems outside of particle physics.

The natural extension of this work is the direct calculation of off-real axis observables through analytic continuation of the SINN amplitude. As each member of a SINN ensemble satisfies all S-matrix principles, constrained errors propagate naturally to all observables in the complex energy plane. A full determination of the calculable quantities is presented in~\cite{letter}. 

Extension to other scattering processes is also possible. $\pi K$ and $\pi N$ scattering are the most direct next tests, as they differ mainly in the form of the dispersion relations required for enforcement of the S-matrix principles. Swapping out the Roy equations for suitable replacements in the relevant channels would immediately allow a full repetition of the analysis presented here.

$\pi\pi$ scattering is a basic input for observables relevant to New Physics searches, such as the hadronic vacuum polarization contribution to $(g-2)_\mu$~\cite{Colangelo:2018mtw,Aoyama:2020ynm} and hadronic final-state interactions relevant for heavier-meson decay analyses~\cite{LHCb:2019jta,Garrote:2022uub}. For $(g-2)_\mu$ input in particular, the same framework can be extended with the corresponding production data and form-factor dispersion relations so that shared systematics and correlations are propagated consistently.

\section*{Data availability}

Each replica in the production SINN ensemble, the medians, and confidence intervals are provided along with an example notebook at Ref.~\cite{wyatt_a_smith_2026_22016314}.

\begin{acknowledgements}
We thank Yaohang Li for his careful reading of the manuscript and suggestions. We thank Emanuele Roberto Nocera and Amedeo Chiefa for helpful discussions on the NNPDF framework. We also thank Marco Filippini and Antonino Fulci for their collaboration in the early stages of this work. This material is based upon work supported by the U.S. Department of Energy, Office of Science, Office of Nuclear Physics under Contract No. 89243126CSC000213, by U.S.~Department of Energy Grant
Nos.~\mbox{DE-FG02-87ER40365}, and \mbox{DE-SC0011090}, and it contributes to the aims of the
U.S.~Department of Energy \mbox{ExoHad} Topical Collaboration, contract \mbox{DE-SC0023598}. ARB acknowledges support by the National Science Foundation (NSF) under Grant No. PHY-2610011. AP and WAS acknowledge support of the GeV-AI project (CUP I57G21000110007), in the context of the ICSC Spoke 2 Open Calls, project funded by European Union -- NextGenerationEU -- and National Recovery and Resilience Plan (NRRP) -- Mission 4 Component 2, and support under the program
``Progetti di Rilevante Interesse Nazionale''
\mbox{(PRIN~2022)}, published on \mbox{2.2.2022} by the Italian Ministry of University and Research (MUR),
Project Title ``The X(3872) files'' -- \mbox{CUP~J53C24002600006} -- Grant Assignment Decree
\mbox{No.~20429} adopted on \mbox{6.11.2024} by the Italian Ministry of University and Research (MUR). This work was supported by the Research Computing clusters at Old Dominion University. The Wahab cluster at Old Dominion University is supported in part by National Science Foundation's grant CNS-1828593.
The authors acknowledge William \& Mary Research Computing for providing computational resources that have contributed to the results reported within this paper.

\end{acknowledgements}

\appendix

\section{Numerical Roy implementation}
\label{app:roy_numerics}

This Appendix records the numerical form of the Roy constraints used in the loss defined in \cref{sec:loss}. For the three constrained channels, $S_0$, $P_1$, and $S_2$, we evaluate the residual
\begin{align}
\Delta_\ell^I(s) \equiv
\mathrm{Re}\, t_\ell^I(s)
- \mathrm{ST}_\ell^I(s)
- \mathrm{KT}_\ell^I(s)
- \mathrm{DT}_\ell^I(s),
\end{align}

on the $N_s=75$ point mesh of \cref{eq:roy-loss}. The subtraction terms $\mathrm{ST}_\ell^I$ are given in  \cref{eq:ST} and the kernel terms in \cref{eq:roy}. The higher-wave set entering the dispersive integrals is $\mathcal{H}=\{D_0,D_2,F_1,G_2\}$.

The high-energy contribution is collected into the driving term. In practice, the fixed-$t$ dispersion relation is split at $\sqrt{s_h}=1.5~\mathrm{GeV}$: the low-energy part is projected into the Roy kernels, while the $s^\prime>s_h$ part is evaluated with fixed Regge input. The corresponding contribution is projected numerically as
\begin{align}
\mathrm{DT}_\ell^I(s)
&=
\frac{1}{64\pi}\int_{-1}^{1} dz_s\, P_\ell(z_s)\,
\mathcal{A}_{\rm Regge}^{I}(s,t(s,z_s)), \label{eq:app_dt} \\
\mathcal{A}_{\rm Regge}^{I}(s,t)
&=
\sum_{I'} \int_{s_h}^{\infty} ds^\prime\,
\mathcal{K}^{I I^\prime}(s,t;s^\prime)\,
\mathrm{Im}\,\mathcal{A}_{\rm Regge}^{I^\prime}(s^\prime,t),
\end{align}
where $\mathcal{K}^{I I^\prime}(s,t;s^\prime)$ are fixed-$t$ dispersive kernels. The \mbox{$s$-channel} Regge amplitudes $\mathcal{A}_{\rm Regge}^{I}$ are built from the \mbox{$t$-channel} Regge amplitudes in Appendix~\ref{app:regge}.

The principal numerical subtlety on the real axis is the principal-value pole of the diagonal kernels at $s'=s$. For the twice-subtracted Roy kernels, the diagonal singular terms have the local structure
\begin{align}
K_{\ell\ell}^{II}(s,s^\prime) =
\frac{N_\ell^I(s,s^\prime)}{s^\prime(s^\prime-s)},
\end{align}
where $N_\ell^I(s,s^\prime)$ is regular at $s^\prime=s$. In the code, $N_\ell^I(s,s^\prime)$ is stored separately from the singular denominator, and the factor $1/s^\prime$ is kept in the numerical denominator. We evaluate the corresponding principal-value integral by subtracting the numerator at the pole,
\begin{align}
&\mathrm{P.V.}\!\int_{4m_\pi^2}^{s_h} ds^\prime\,
\frac{f(s^\prime,s)}{s^\prime(s^\prime-s)}
=
\int_{4m_\pi^2}^{s_h} ds^\prime\,
\frac{f(s^\prime,s)-f(s,s)}{s^\prime(s^\prime-s)}
\nonumber\\
&\quad
{}+ f(s,s)\,
\mathrm{P.V.}\!\int_{4m_\pi^2}^{s_h}\frac{ds^\prime}{s^\prime(s^\prime-s)},
\end{align}
with $f(s^\prime,s)=N_\ell^I(s,s^\prime)\,\mathrm{Im}\,t_\ell^I(s^\prime)$.The first integral is regular and is evaluated by Gauss-Legendre quadrature over fixed subdomains; the second is performed analytically,
\begin{align}
\mathrm{P.V.}\!\int_{4m_\pi^2}^{s_h}\frac{ds^\prime}{s^\prime(s^\prime-s)}
=\frac{1}{s}\log\left[\frac{4m_\pi^2}{s_h}\left|\frac{s_h-s}{s-4m_\pi^2}\right|\right].
\end{align}
Once $s$ is continued off the real axis, the pole is displaced from the integration contour and the same quadrature can be performed without a principal-value subtraction.

\section{Regge physics}
\label{app:regge}

This Appendix specifies the fixed Regge input used above the matching point $\sqrt{s_h}=1.5~\mathrm{GeV}$, following the standard dispersive practice~\cite{Pelaez:2003ky,Yndurain:2007qm,GarciaMartin:2011jx,Caprini:2011ky}. At high energy, a crossed-channel Regge representation replaces the slowly convergent finite $s$-channel partial wave expansion. We use the Regge description of Refs.~\cite{Pelaez:2003ky,GarciaMartin:2011cn}, with all parameters held fixed. Because the dispersion relations are twice subtracted, the high-energy input is strongly suppressed in the low-energy region. Other Regge descriptions, including those of Ref.~\cite{Caprini:2011ky}, give equivalent low-energy driving terms at the precision relevant here.

This Regge input is naturally written in terms of \mbox{$t$-channel} isospin amplitudes $\mathcal{A}^{(I_t)}$, denoted by parentheses. The \mbox{$s$-channel} amplitudes entering the Roy driving terms are
\begin{equation}
\begin{aligned}
\mathcal{A}_{\rm Regge}^0(s,t) &=
\frac{1}{3}\mathcal{A}^{(0)}(s,t)
+\mathcal{A}^{(1)}(s,t)
+\frac{5}{3}\mathcal{A}^{(2)}(s,t), \\
\mathcal{A}_{\rm Regge}^1(s,t) &=
\frac{1}{3}\mathcal{A}^{(0)}(s,t)
+\frac{1}{2}\mathcal{A}^{(1)}(s,t)
-\frac{5}{6}\mathcal{A}^{(2)}(s,t), \\
\mathcal{A}_{\rm Regge}^2(s,t) &=
\frac{1}{3}\mathcal{A}^{(0)}(s,t)
-\frac{1}{2}\mathcal{A}^{(1)}(s,t)
+\frac{1}{6}\mathcal{A}^{(2)}(s,t).
\end{aligned}
\end{equation}

Only the imaginary parts of the Regge amplitudes enter the dispersive integrals used here. The $I_t=0$ amplitude is described in terms of the Pomeron and an additional $P'$ exchange corresponding to the $f_2(1270)$ resonance. The imaginary part of the Regge amplitude is
\begin{align}
&\mathrm{Im}\,\mathcal{A}^{(0)}(s,t) = P(s,t) + P^\prime(s,t),  \nonumber \\
&P(s,t) =
\beta_P \Psi_P(t)\,\alpha_P(t)\,\frac{1+\alpha_P(t)}{2}\,
e^{bt}\left(\frac{s}{s_0}\right)^{\alpha_P(t)}, \nonumber\\
&P^\prime(s,t) =
\beta_{P^\prime} \Psi_{P^\prime}(t)\,
\frac{\alpha_{P^\prime}(t)\,[1+\alpha_{P^\prime}(t)]}
{\alpha_{P^\prime}(0)\,[1+\alpha_{P^\prime}(0)]}\,
e^{bt}\left(\frac{s}{s_0}\right)^{\alpha_{P^\prime}(t)}.
\end{align}

For $I_t=1$ we use $\rho$ exchange,
\begin{align}
\mathrm{Im}\,\mathcal{A}^{(1)}(s,t) &=
\beta_\rho \frac{1+\alpha_\rho(t)}{1+\alpha_\rho(0)}
\varphi_\rho(t)\,e^{bt}\left(\frac{s}{s_0}\right)^{\alpha_\rho(t)},
\end{align}
and for $I_t=2$ we use a small effective exchange,
\begin{align}
\mathrm{Im}\,\mathcal{A}^{(2)}(s,t) =
\beta_2\,e^{bt}\left(\frac{s}{s_0}\right)^{\alpha_\rho(t)+\alpha_\rho(0)-1}.
\end{align}

All $t$-dependence is measured in $\mathrm{GeV}^2$, and all Regge parameters are held fixed. Exact details and values for the parameters were provided by the authors of Ref.~\cite{GarciaMartin:2011cn}.

The fixed Regge amplitudes enter the $s'>s_h$ contribution to the driving term defined in \cref{eq:app_dt}. The high-energy contribution is smooth and strongly suppressed at low energy by the twice-subtracted Roy kernels. It supplies the fixed asymptotic completion required by the dispersion relation.

\section{Spectral clustering and group rejection}
\label{app:selection_eigenspace}

\begin{figure*}[!ht]
\centering
\includegraphics[width=0.75\linewidth]{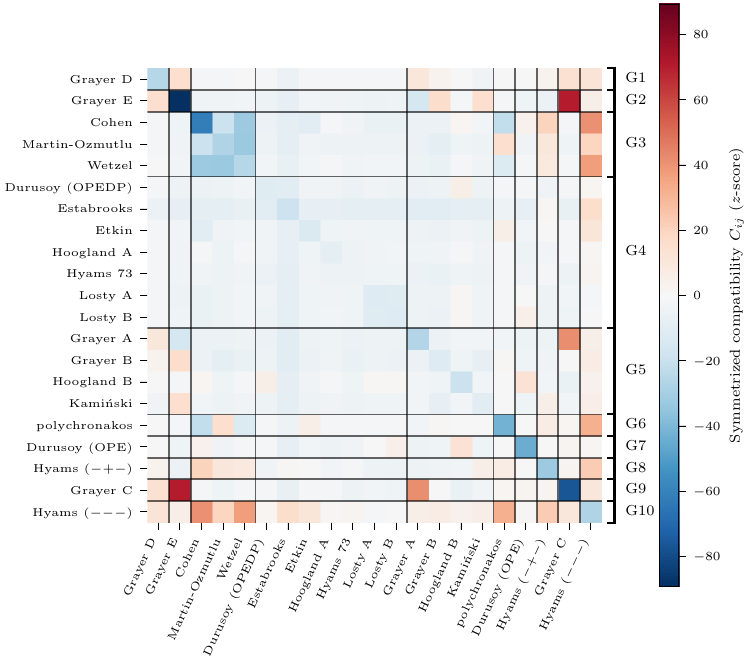}
\caption{Experiment-level symmetrized response matrix $C_{ij}$ for the second selection round, ordered by Ward group. Positive entries indicate tension and negative entries indicate compatibility; the group-averaged criterion and rejection threshold are described in \cref{subsec:selection_strat}.}
\label{fig:full_response}
\end{figure*}

This Appendix records the specifics of the second spectral response clustering round described in~\cref{subsec:retained}.

\begin{figure*}[!ht]
\centering
\includegraphics[width=0.75\linewidth]{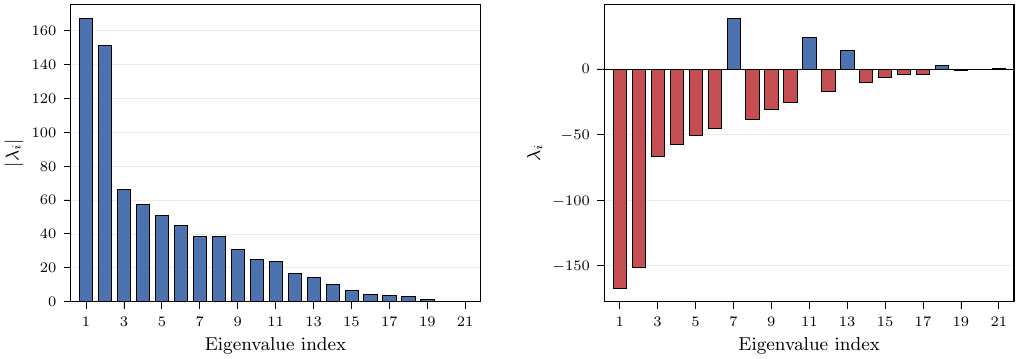}
\caption{Eigenvalue spectrum of the response matrix $C_{ij}$. The first 11 eigenvalues capture 90\% of the total absolute spectral weight and are retained for the clustering.}
\label{fig:eigenvalues}
\end{figure*}

After the first round, the surviving global data set contained $23$ experiments, two of which were retained, leaving $21$ experiments for the second spectral response clustering round. Its response matrix is shown in~\cref{fig:full_response}. Applying the spectral decomposition~\cref{eq:response-eig} to this matrix and retaining eigenvectors until the cumulative absolute spectral weight reaches the $90\%$ level selects \mbox{$k=11$} modes. The ordered eigenvalue spectrum is shown in~\cref{fig:eigenvalues}.

\begin{figure}[!ht]
\centering
\includegraphics[width=\linewidth]{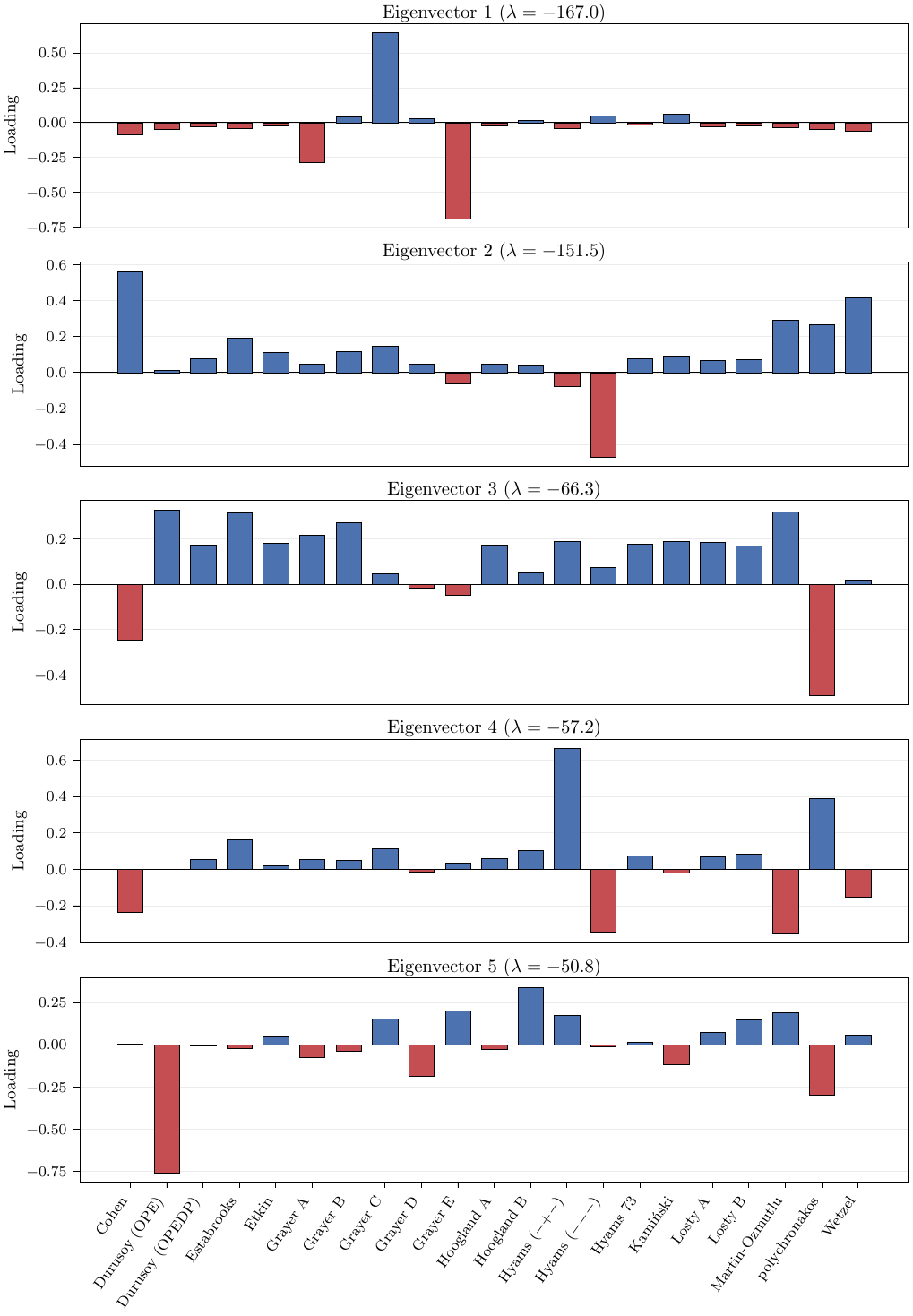}
\caption{Leading eigenvector loadings for each experiment. Experiments with similar loading patterns are grouped together by the Ward clustering.}
\label{fig:eigenvectors}
\end{figure}

The eigenvector loadings in \cref{fig:eigenvectors} identify experiments that share similar response patterns before the final pruning step. Experiments with the same sign and magnitude pattern across the leading modes are merged early in the Ward hierarchy, whereas experiments with distinct loadings remain separated until higher merge heights.
 
Ward clustering on the embedded experiments produces the dendrogram, which is cut at the largest merge-height gap to yield $n_{\mathrm{g}}=10$ candidate groups, shown in~\cref{fig:dendrogram} and listed in~\cref{tab:ward-groups}. Seven of these groups contain a single experiment; the remaining three are multi-member clusters.

\begin{table}[!ht]
\caption{Ward groups for the second selection round before the rejection step.}
\label{tab:ward-groups}
\centering
\small
\begin{tabular}{c l}
\toprule
Group & Members \\
\midrule
G1 & Grayer D \\
G2 & Grayer E \\
G3 & Cohen, Martin-Ozmutlu, Wetzel \\
G4 & \parbox[t]{0.66\columnwidth}{$\!\!\!\!\!\!\!$Durusoy OPEDP, Estabrooks, Etkin,
Hoogland A, Hyams 73, Losty (A,B)} \\
G5 & Grayer (A,B), Hoogland B, Kami\'{n}ski \\
G6 & Polychronakos \\
G7 & Durusoy OPE \\
G8 & Hyams $(-\!+\!-)$ \\
G9 & Grayer C \\
G10 & Hyams $(-\!-\!-)$ \\
\bottomrule
\end{tabular}
\end{table}

At each iteration of the group removal defined in~\cref{subsec:selection_strat}, every still-active group $G_\ell$ is assigned the list of partners $G_m$ for which $\bar{C}_{\ell,m} > +1$. The removed group is the one participating in the largest number of such violations. Ties are broken first by the largest violated entry and then by removing the smaller group. The procedure does not recompute the Ward partition after a removal; each group keeps the membership assigned by the dendrogram cut.

The five removal steps are:
\begin{enumerate}
\item \textbf{Remove G10} (Hyams $(-\!-\!-)$): twenty between-group entries exceed $+1$, with a maximum group-averaged response $\bar{C}_{10,3}=+32.2$ against G3.
\item \textbf{Remove G9} (Grayer C): twelve surviving pairs still exceed $+1$, with maximum $\bar{C}_{9,2}=+69.9$ against G2.
\item \textbf{Remove G1} (Grayer D): six violations above $+2.5$ remain, with maximum $\bar{C}_{1,2}=+15.0$ against G2.
\item \textbf{Remove G8} (Hyams $(-\!+\!-)$): three violations above $+1$ remain, with maximum $\bar{C}_{8,3}=+12.8$ against G3.
\item \textbf{Remove G2} (Grayer E): one violation above $+1$ remains, $\bar{C}_{2,5}=+3.5$ against G5.
\end{enumerate}
After these five steps, the surviving groups G3, G4, G5, G6, and G7 contain the $16$ clustered experiments retained by the procedure, and the maximum surviving between-group average is $+0.3$.

\section{Auxiliary partial waves}
\label{app:supp_results}

\begin{figure*}[t]
\centering
\includegraphics[width=0.95\linewidth]{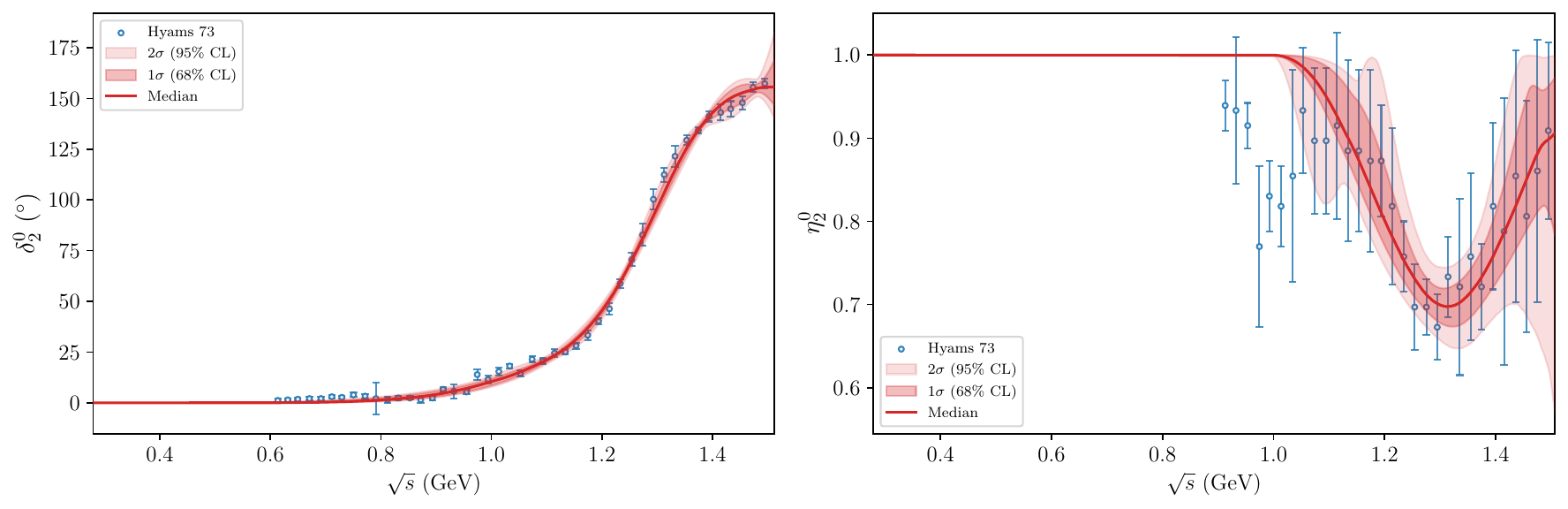}
\caption{$D_0$ phase shift and inelasticity. Bands and data as in \cref{fig:waves_main}.}
\label{fig:D0}
\end{figure*}

\begin{figure*}[t]
\centering
\includegraphics[width=0.95\linewidth]{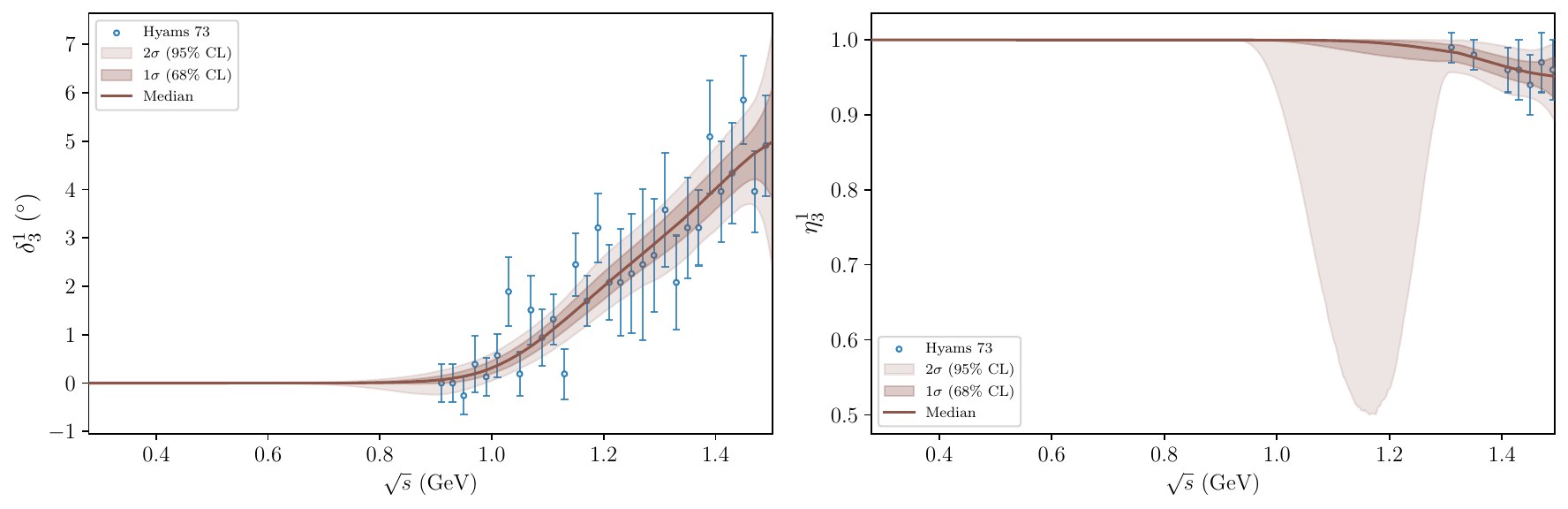}
\caption{$F_1$ phase shift and inelasticity. Bands and data as in \cref{fig:waves_main}. The $F_1$ partial is underspecified due to poor data coverage above threshold, and also underconstrained as it is not controlled by the Roy equations. The large fluctuation in the $95\%$ uncertainty band is an artifact of the truncation of the dispersive integral.}
\label{fig:F1}
\end{figure*}

\begin{figure}[t]
\centering
\includegraphics[width=\linewidth]{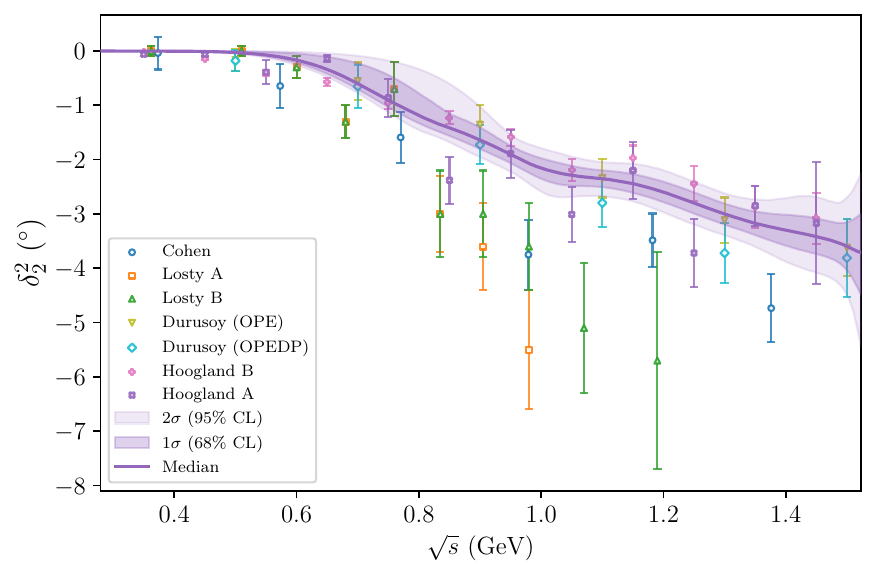}
\caption{$D_2$ phase shift (elastic).}
\label{fig:D2}
\end{figure}

\begin{figure}[t]
\centering
\includegraphics[width=\linewidth]{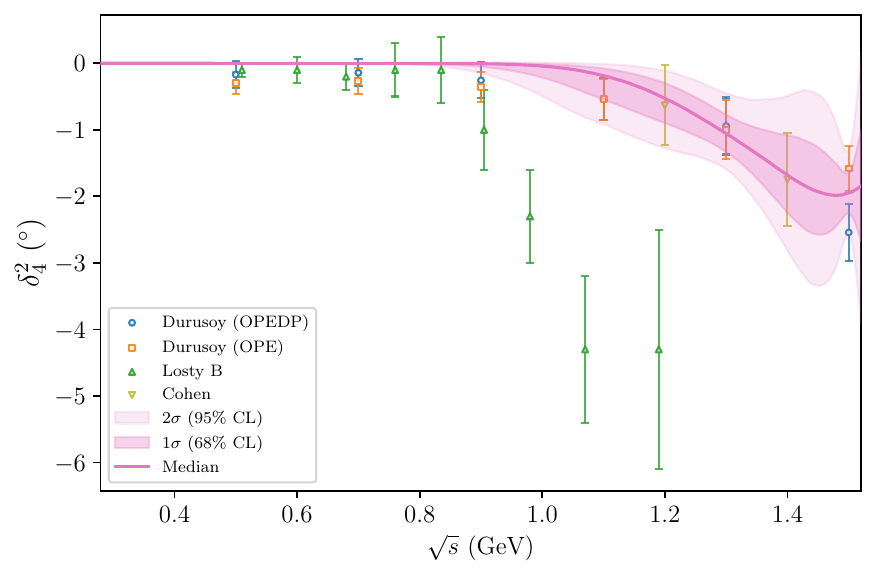}
\caption{$G_2$ phase shift (elastic).}
\label{fig:G2}
\end{figure}

As explained in the main text, the $D_0$, $F_1$, $D_2$, and $G_2$ partial waves are required as input to the Roy system, since their contributions remain non-negligible up to $s_h$. In principle, dispersive constraints could also be imposed on them~\cite{Bydzovsky:2016vdx}. In the present analysis, however, they play an auxiliary role: they are needed to complete the dispersive input, but our primary physics targets lie in the $S_0$, $P_1$, and $S_2$ channels. \Cref{fig:D0,fig:F1,fig:D2,fig:G2} show the resulting descriptions. 

\section{Training diagnostics}
\label{app:supp_validation}

\begin{figure*}[t]
\centering
\includegraphics[width=0.95\linewidth]{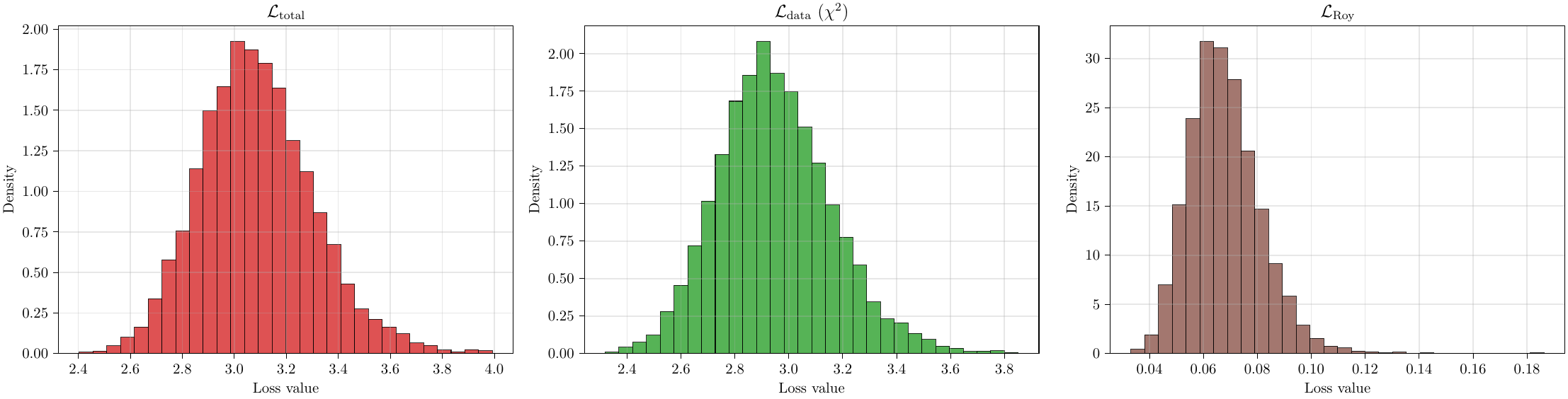}
\caption{Loss components across the $10{,}000$-replica production ensemble, every member satisfying \mbox{$\mathcal{L}_{\text{total}}\le4.0$}.}
\label{fig:losses}
\end{figure*}

\begin{figure*}[t]
\centering
\includegraphics[width=0.95\linewidth]{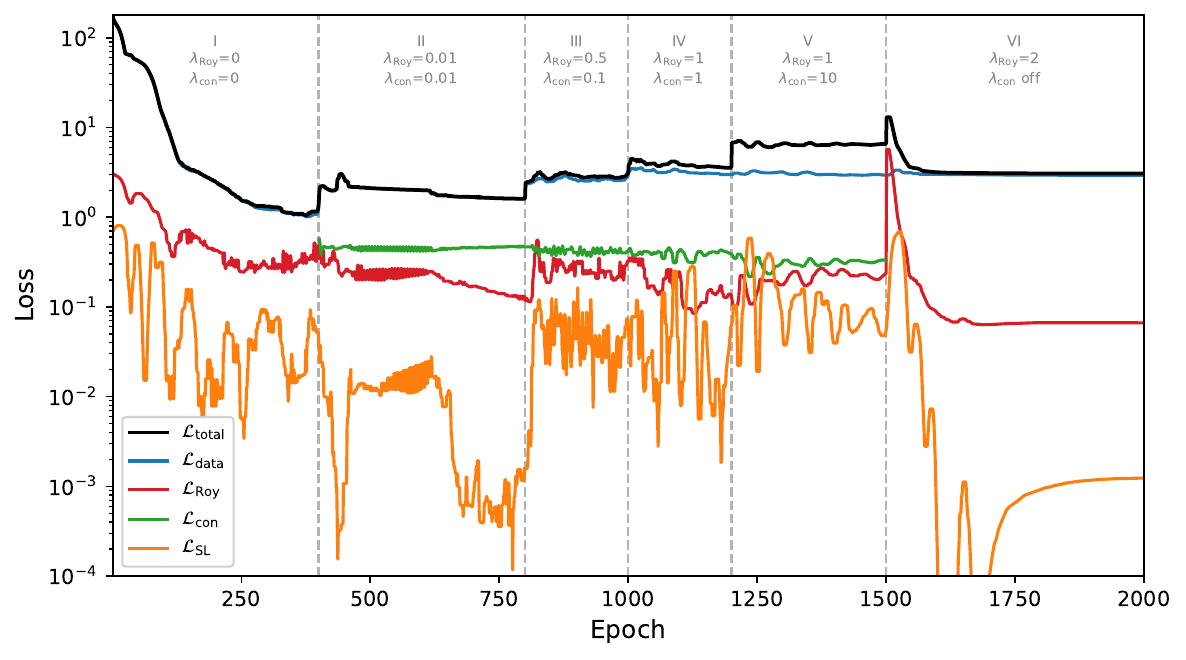}
\caption{Training loss evolution for a representative replica: total (black), data (blue), Roy (red), consistency (green), and DIRAC (orange) terms of~\cref{eq:total-loss}. $\mathcal{L}_{\mathrm{SL}}$ carries its fixed $1/N_\mathrm{data}$ normalization; the other components are shown unweighted and only the total carries the curriculum weights, which are annotated per phase.}
\label{fig:training}
\end{figure*}

\begin{figure}[t]
\centering
\includegraphics[width=0.9\linewidth]{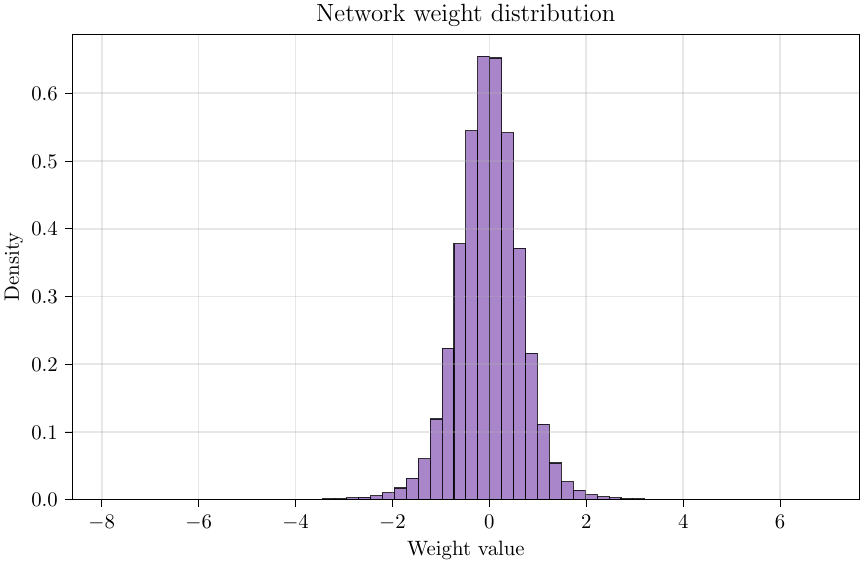}
\caption{Distribution of trained network weights across the retained ensemble. The peak near zero indicates substantial effective sparsity.}
\label{fig:weights}
\end{figure}

\begin{enumerate}
\item \textit{Production loss.} \Cref{fig:losses} shows the histograms of the loss distributions for the production ensemble. All loss components converge well across the ensemble, and exhibit approximately Gaussian shapes.
\item \textit{Training convergence.} \Cref{fig:training} shows the loss evolution for a representative training run across the six curriculum phases. The spikes at phase boundaries are expected: they coincide with the activation of additional loss terms. When the Roy penalty is introduced, $\mathcal{L}_{\mathrm{Roy}}$ rises briefly while the network readjusts from the data-only solution, after which it relaxes quickly inside the new phase. The same pattern appears when the scattering-length consistency term is turned on. In the representative run shown, the retained loss components stabilize during the final phase.
\item \textit{Overparametrization.} The production network carries $1{,}156$ trainable weights against $621$ retained data points, so the fit is nominally overparametrized. \Cref{fig:weights} shows the trained weight distribution across the retained ensemble, peaked near zero. This confirms that the weight decay is active, but it is not on its own evidence against overfitting, since decoupled weight decay produces a zero-peaked histogram in any case. The imposed physics collapses the space of admissible functions far below the raw parameter count: the threshold factors and the inelasticity map fix the form of every output exactly, and the coupled Roy residuals tie the seven waves to each other and to the subtraction constants. Whatever freedom survives is measured directly, by the closure test of~\cref{sec:validation_closure}, which recovers a known target in every threshold observable, and by the replica spread, which is the variance of independently retrained networks over resampled data.
\end{enumerate}

\section{Numerical implementation}
\label{app:hyperparameters}

This Appendix collects the numerical settings and final implementation choices that define the production SINN runs. The architectural motivation, hybridization layer, curriculum logic, and Roy constraints are discussed in the main text and the preceding Appendices.

\begin{table}[!htb]
\caption{Network layer dimensions. Numbers are hidden neurons before the final single-neuron output. The head subnetworks use $\mathrm{SiLU}$ activations, while the shared trunk uses the fixed activation pattern discussed below.}
\label{tab:architecture}
\centering
\begin{tabular}{l c c c}
\toprule
Component & $\ell$, $I$ & Phase layers & Inel.\ layers \\
\midrule
Trunk & --- & [8, 12] & --- \\
\midrule
$S_0$ head & 0, 0 & [8, 4] & [4] \\
$P_1$ head & 1, 1 & [4, 2] & [2] \\
$S_2$ head & 0, 2 & [4, 2] & [2] \\
$D_0$ head & 2, 0 & [2, 2] & [2] \\
$D_2$ head & 2, 2 & [4, 2] & --- \\
$F_1$ head & 3, 1 & [4, 2] & [2] \\
$G_2$ head & 3, 2 & [4, 2] & --- \\
\bottomrule
\end{tabular}
\end{table}

\begin{enumerate}
\item \textit{Architecture and output map.}
Beyond the neuron counts listed in~\cref{tab:architecture}, the production runs fix the shared trunk to the activation pattern $[\tanh,\mathrm{SiLU}]$, and use learned residual projections between trunk blocks. This compact trunk-plus-head layout is among the smallest we found that can still describe the full data set while satisfying the coupled Roy constraints. Head sizes are assigned by wave complexity: the data-rich $S_0$ channel uses the largest head, whereas the smaller $D_2$ and $G_2$ channels use minimal heads. The $\tanh$ first block helps bound the internal representation during the early data-dominated stages, and the residual projections improve the transmission of the Roy-driven gradient back through the shared trunk. 
 
The exact threshold factors and exponential inelasticity map are part of the postprocessing layer described in~\cref{sec:architecture}; the production-specific inelastic-channel choices are listed in~\cref{tab:inel_channels}.

\item \textit{Optimizer, schedule, and stopping.}
The numerical optimization settings not stated elsewhere are collected in~\cref{tab:optimization_settings}. The production runs use the AdamW, UPGrad, and curriculum strategy described in~\cref{sec:loss,sec:curriculum}. AdamW has learning rate $\alpha=0.005$ and weight decay $\lambda_{\mathrm{wd}}=0.001$ is used throughout. Higher rates destabilize the Roy penalty in the middle curriculum phases, while lower rates slow convergence in the data-only phase. Decoupled weight decay is useful here because the multi-objective solver modifies gradients before the optimizer step. Gradient clipping at norm $\le 1.0$ prevents large spikes when the amplitude is still minimally Roy-constrained. Cosine annealing with step size 100 epochs, decay factor 0.5, and minimum learning rate $\alpha_{\min}=10^{-5}$ is used in production. A patience criterion of $50$ epochs with $\delta_{\min}=10^{-6}$ is applied to the total training objective. In intermediate curriculum phases, triggering early stopping advances to the next phase and resets the counter; in the final phase it terminates training.

\item \textit{Initialization and auxiliary numerical settings.}
All weights are initialized with Xavier-normal draws and gain 1.4. The network scattering lengths entering the final Roy subtraction constants are computed from a four-point Richardson extrapolation to threshold with step $h=10^{-4}~\mathrm{GeV}^2$.
The experimental data are a few decades old, and have been collected from various sources. While we group the data by different manuscripts, some of these come from reanalyses or updates, and some data points might be repeated. Exact duplicate points are removed with tolerance $10^{-5}$ during data set assembly to avoid double counting. 

\item \textit{Variants not retained.}
Several choices were explored during development but were not selected for the final runs. In postprocessing, $p\cot\delta$ trigonometric maps were tested to describe Breit-Wigner-like lineshapes. While they produced promising results in most of the energy region, they became numerically unstable near $\delta=180^\circ$ in the $P_1$ and $D_0$ waves. Numerical barrier-factor capping and additional phase or inelasticity smoothing did not provide an advantage over the hybridized threshold and inelasticity maps already used in production. Likewise, a sigmoid inelasticity map was not retained once the exponential form with the chosen threshold onset was implemented. Adaptive loss reweighting, explicit per-channel data balancing, data dropout, smooth curriculum transitions, and early-abort checks for poor initializations did not improve the final constrained fits. Alternative optimizers and learning-rate schedules were also tested, but none outperformed the simpler AdamW-plus-cosine setup used here.
\end{enumerate}

\begin{table}[!htb]
\caption{Optimization settings used in production that are not specified explicitly elsewhere in the manuscript.}
\label{tab:optimization_settings}
\centering
\begin{tabular}{l c}
\toprule
Optimization setting & Value \\
\midrule
Learning rate & $0.005$ \\
Weight decay & $0.001$ \\
$(\beta_1,\beta_2,\epsilon)$ & $(0.9,0.999,10^{-8})$ \\
Gradient clipping & norm $\le 1.0$ \\
LR scheduler & cosine annealing \\
Scheduler step / factor & 100 epochs / 0.5 \\
Minimum learning rate & $10^{-5}$ \\
Early-stopping $\delta_{\min}$ & $10^{-6}$ \\
Weight initialization & Xavier normal, gain 1.4 \\
\bottomrule
\end{tabular}
\end{table}

\begin{figure*}[t]
\centering
\includegraphics[width=0.92\linewidth]{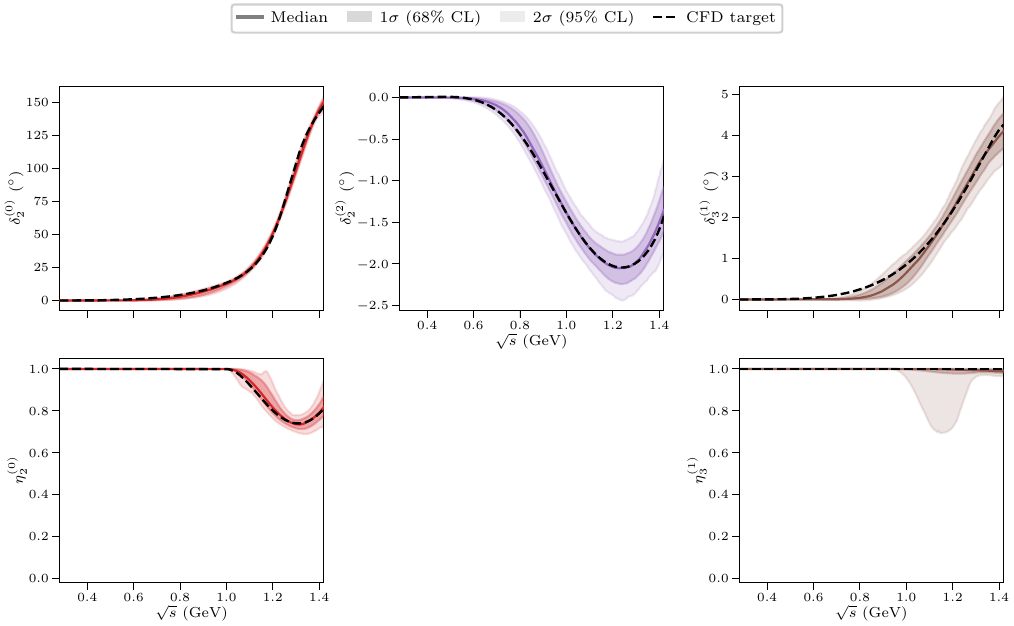}
\caption{As \cref{fig:closure_lineshapes_low}, for the $D_0$, $D_2$, and $F_1$ waves.}
\label{fig:closure_lineshapes_high}
\end{figure*}


\section{Architecture sweep protocol}
\label{app:sweep_protocol}

The $500$-trial sweep of \cref{sec:validation_architecture} was run with Ray Tune~\cite{liaw2018tune},\footnote{\url{https://docs.ray.io/en/latest/tune/}} over the following $21$ hyperparameter dimensions, organized into four groups:
\begin{enumerate}
\item \textit{Trunk geometry}: number of hidden layers, neurons per layer, and activation functions;
\item \textit{Per-channel head structure}: for each of the 7 channels independently, the head depth and activation functions;
\item \textit{Physics switches}: The effective $S_2$ inelastic threshold mass sampled, $m_{\mathrm{eff}} \in [0.45, 0.65]~\mathrm{GeV}$;
\item \textit{Optimizer and constraint weights}: learning rate ($10^{-4}$--$2{\times}10^{-2}$, log-uniform sampled), weight decay ($10^{-5}$--$5{\times}10^{-3}$, log-uniform sampled), and the final-phase Roy weight $\lambda_{\mathrm{Roy}}$ ($0.25$--$8$, log-uniform sampled).
\end{enumerate}

The bulk of the variation is in the per-channel heads: 14 of the 21 dimensions explored in our hyperparameter sweep control the independent structure, nonlinearity, and capacity of each channel's output network. Across the sweep, total parameter counts range from $414$ to $4{,}266$ (median $1{,}649$), compared with $1{,}156$ for the production architecture. Each configuration was trained on $10$ independent Gaussian data replicas, drawn exactly as in~\cref{sec:ensemble}, using the same curriculum, loss function, and early stopping as the production ensemble.

All activation functions tested were smooth. Piecewise-linear activations such as $\mathrm{ReLU}$ interfere with the smoothness of the partial waves, which follows from analyticity (and causality), and can therefore spoil fulfillment of the Roy equations.

\section{Closure-test construction}
\label{app:closure}

This Appendix records the construction of the closure test whose results are summarized in~\cref{sec:validation_closure}.

\begin{figure}[b]
\centering
\includegraphics[width=0.95\linewidth]{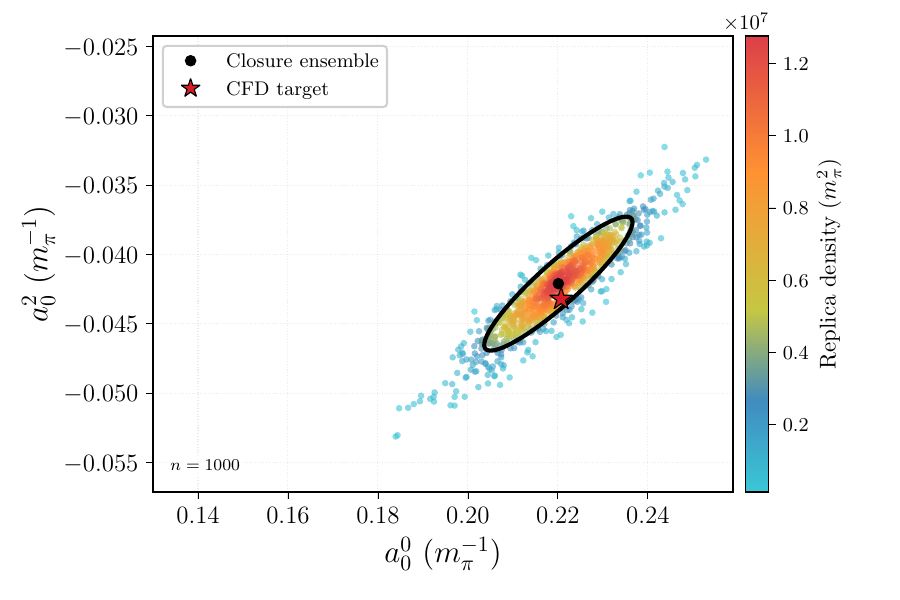}
\caption{Closure test, joint distribution of $a_0^0$ and $a_0^2$ over the CFD ensemble, with the CFD target marked.}
\label{fig:closure_sl}
\end{figure}

\begin{enumerate}
\item \textit{Pseudo-data generation.}
Six of the seven channels take their central values from the CFD parametrization~\cite{GarciaMartin:2011cn}; the CFD fit does not include the $G_2$ partial wave, and so its data are replaced by zero for phase shifts and one for inelasticities.
\item \textit{Training.}
Training follows the production procedure exactly, with each replica drawing one independent Gaussian fluctuation about the CFD central values as in~\cref{sec:ensemble}. Each of the $1{,}000$ retained replicas satisfies a total-loss cut $\mathcal L_\mathrm{total}\leq4.5$.
\item \textit{Real-axis reconstruction.}
\Cref{fig:closure_lineshapes_high} compares the reconstructed $D_0$, $D_2$, and $F_1$ waves with the CFD amplitude that generated the data. The agreement is as good as for the constrained waves of~\cref{fig:closure_lineshapes_low}, and the bands widen only where the pseudo-data inherit sparse or imprecise coverage from the real data set.
\item \textit{Derived quantities.}
\Cref{fig:closure_sl} shows the joint distribution of the scattering lengths, where the target lies close to the center of the $68\%$ ellipse and the reconstructed $a_0^0$--$a_0^2$ correlation reproduces that of the generating amplitude.
\end{enumerate}

\bibliographystyle{apsrev4-2}
\bibliography{bibliography}
\end{document}